\documentclass[envcountsame]{llncs}
\usepackage[utf8]{inputenc}
\usepackage[T1]{fontenc}
\usepackage[hidelinks]{hyperref}
\usepackage{amsmath}
\usepackage{amssymb}
\usepackage{underscore} 
\usepackage{microtype}
\usepackage{comment}
\usepackage{subfigure}
\usepackage{tikz}
\usepackage{todonotes}
\usetikzlibrary{shapes.geometric}
\usetikzlibrary{calc}
\usetikzlibrary{fit}
\usetikzlibrary{patterns}
\usetikzlibrary{decorations.pathreplacing}
\makeatletter\def\@LN#1#2{}\makeatother

\makeatletter
\def\@doletters#1{\def\@b{#1}\ifx\@b\@c\let\@a\egroup\else\@d{#1}\fi\@a}
\def\doletters#1#2{\bgroup\def\@d##1{\global\@namedef{#1##1}{\ensuremath{#2{##1}}}}\def\@a{\@doletters}\def\@c{;}\@a}
\doletters c\mathcal  ABCDEFGHIJKLMNOPQRSTUVWXYZ;
\doletters s\mathsf   ABCDEFGHIJKLMNOPQRSTUVWXYZ
                    {BW}{BL}{LD}{DL}{RSA}{NFS}{SNFS}{GNFS};
\doletters b\mathbb   ABCDEFGHIJKLMNOPQRSTUVWXYZ;
\doletters {fk}\mathfrak ABCDEFGHIJKLMNOPQRSTUVWXYZ
                         abcdefghijklmnopqrstuvwxyz;
\doletters {bf}\mathbf   ABCDEFGHIJKLMNOPQRSTUVWXYZ
                         abcdefghijklmnopqrstuvwxyz;

\newcommand{\rquo}[2]{\leavevmode\kern-.1em\raise.2ex\hbox{$#1$}\kern-.1em/\kern-.1em\lower.25ex\hbox{$#2$}}
\newcommand{\lquo}[2]{\leavevmode\kern-.1em\lower.25ex\hbox{$#2$}\kern-.1em\backslash\kern-.1em\raise.2ex\hbox{$#1$}}
\newcommand{\GF}[1]{\texorpdfstring{\ensuremath{\bF_{#1}}}{GF(#1)}}
\newcommand{\ZnZ}[1]{\texorpdfstring{\ensuremath{\rquo{\bZ}{#1\bZ}}}{Z/#1Z}}
\DeclareMathOperator{\Norm}{Norm}
\DeclareMathOperator{\Res}{Res}
\def\cadonfs{Cado-NFS}
\makeatother

\newcommand{\ablue}{blue!50!lightgray}
\newcommand{\aorange}{red!50!yellow}
\newcommand{\ared}{red!50!lightgray}

\usepackage{footnote}
\makesavenoteenv{tabular}
\makesavenoteenv{table}

\def\stackfrac{\genfrac{}{}{0pt}{}}

\newif\iffinal\finaltrue

\begin{document}

\title{Comparing the difficulty of factorization and discrete logarithm:
  a 240-digit experiment%
\thanks{\textcopyright IACR 2020. This article is the final version submitted by the
  authors to the IACR and to Springer-Verlag on 2020-06-09 for CRYPTO2020,
  available on ePrint at \url{https://eprint.iacr.org/2020/xxxx}.
  The version published by Springer-Verlag will be available in August 2020.}
}

\iffinal
\author{Fabrice~Boudot\inst{1} \and
        Pierrick~Gaudry\inst{2} \and
        Aurore~Guillevic\inst{2} \and
        Nadia~Heninger\inst{3} \and
        Emmanuel~Thomé\inst{2} \and
        Paul~Zimmermann\inst{2}}
\institute{%
    Université de Limoges, XLIM, UMR 7252, F-87000 Limoges, France
    \and
    Université de Lorraine, CNRS, Inria, LORIA, F-54000 Nancy, France
    \and
    University of California, San Diego, USA\\
    \vbox to 0pt{\vskip 6pt{\em In memory of Peter L.~Montgomery}\vss}}
\fi


\maketitle
\begin{abstract}
  We report on two new records:
  the factorization of RSA-240, a 795-bit number,
  and a discrete logarithm computation over a 795-bit prime field.
  Previous records were the factorization of RSA-768 in 2009 and
    a 768-bit discrete logarithm computation in 2016.
  Our two computations at the 795-bit level were done using the same hardware and software,
  and show that computing a discrete logarithm is not much harder than
  a factorization of the same size. Moreover, thanks to algorithmic
    variants and well-chosen parameters, our computations were significantly
    less expensive than anticipated based on previous records.

  The last page of this paper also reports on the factorization of
    RSA-250.
\end{abstract}

\section{Introduction}

The Diffie-Hellman protocol over finite fields and the RSA cryptosystem
were the first practical building blocks of public-key cryptography.
Since then, several other cryptographic primitives have entered the
landscape, and a significant amount of research has been put into the development,
standardization, cryptanalysis, and optimization of
implementations for a large number of cryptographic primitives. Yet the
prevalence of RSA and finite field Diffie-Hellman is still a fact:
between
November 11, 2019 and December 11, 2019, 
the ICSI Certificate Notary~\cite{icsisslnotary} observed that 
90\% of the TLS certificates used RSA signatures,
and 7\% of the TLS connections used RSA for key exchange.
This holds despite the
much longer key sizes required by these primitives compared to elliptic curves:
based on the asymptotic formulas for the best known cryptanalysis algorithms, 
the required key size for RSA or finite field Diffie-Hellman is roughly estimated 
to grow as a cubic function of the security
parameter, while the required key size for elliptic curve
cryptosystems grows only as a linear function of the security
parameter.\footnote{A security parameter $\epsilon$ asserts that
cryptanalysis requires $2^\epsilon$ operations; assuming Moore's law, the
security parameter could be seen as growing linearly with time.}

Over the last few years, the threat posed by quantum computers has been used as a justification to
postpone the impending replacement of
RSA and finite field Diffie-Hellman by alternatives such as elliptic
curve cryptography~\cite{nsa-thing-2015}, resulting in implementation choices that seem paradoxical from the perspective of classical cryptanalysis.

Key sizes for RSA and finite field Diffie-Hellman have become unwieldy. To
meet a 128-bit security strength, it is widely accepted that both schemes require a key size
of approximately 3072 bits (see e.g., the 2018 ECRYPT-CS Recommendations).
While it is easy to deal with
such key sizes in environments where computing power is plentiful (laptop
and desktop computers, or cell phones), a surprising amount of public key cryptography in use employs
weak key strengths. There are two main factors contributing to the use of weak key sizes in practice. First, implementations may use weak key sizes to ensure backward compatibility.  For example, a crucial component in the FREAK and Logjam attacks
was the widespread support for weakened ``export-grade'' cipher suites using 512-bit keys~\cite{logjam}; the Java JDK versions 5-8 originally supported Diffie-Hellman and DSA primes of at most 1024 bits by default.
Second, in embedded environments, or
when very little computational power is allotted to public-key
cryptographic operations, small key sizes are not rare. As an example, in
2018, an off-the-shelf managed network switch purchased by the authors
shipped with a default RSA ssh host key of 768 bits (despite a \$2,000 USD
list price), a key size that has been
broken since 2009 in the academic world.

The main goal of this article is to assess the difficulty of the
mathematical problems that underpin the security of RSA and finite field
Diffie-Hellman and DSA, namely
integer factorization (IF) and discrete logarithm (DL). We are
interested both in the feasibility of cryptanalysis of these problems as well
as in their relative difficulty.
Our targets are RSA-240, from the RSA Factoring Challenge,
and DLP-240, denoting the challenge of computing discrete logarithms
modulo $p = \text{RSA-240} + 49204$, which is the smallest safe prime above
RSA-240 (i.e., 
$(p-1)/2$ is also prime).
Both previous records were 768-bit keys, with results dating back to
2009~\cite{C:KAFLTB10} and 2016~\cite{EC:KDLPS17}.
The total cost of our computation is about 1000 core-years for RSA-240,
and 3200 core-years for DLP-240. Here and throughout this article, 
the core-years we mention are relative to the computing platform that we
used most, namely
Intel Xeon Gold 6130 CPUs with 16 physical cores
(32 hyperthreaded cores) running at 2.10GHz. A core-year is the use of
one of these physical cores for a duration of one year.
As in the previous records, our computations used the Number Field Sieve
algorithm (NFS for short),
which has variants both for integer factoring and finite field
discrete logarithms.

Improvements in cryptanalysis records are to be expected.
In this article, our contribution
is not limited to reaching new
milestones (and reminding people to get rid of outdated keys).
Rather, it is interesting to report on \emph{how} we reached them:

\begin{itemize}
\item We
developed a testing framework that enabled us to precisely select,
among a wide variety with complex interactions, parameters that influence the running
time of NFS. We were able to accurately predict important
measures such as the matrix size.

\item Some folklore ideas that have been known for some time in the
    NFS community played a very important role in our computation.
In particular the \emph{composite special-\fkq} used in
relation collection for DLP-240 proved extremely beneficial, and
        so did \emph{batch smoothness detection}, which we used both for
        RSA-240 and DLP-240.
This is the first time that this latter technique has been used in a factoring
        record for general numbers (it was used in~\cite{AC:KleBosLen14},
        in a very specific context). These techniques, together with our careful parameter selection,
contributed to a significantly faster computation
than extrapolation from the running times of
previous records would have suggested. Even on similar hardware, our relation collection
effort for the 795-bit DLP computation took 25\% less time than the
reported relation collection effort of the previous 768-bit DLP record.

\item Furthermore, we computed two records of the same
size, RSA-240 and DLP-240, at the same time and using 
hardware of the same generation. This is completely new and gives a 
crucial data point for the
assessment of the relative difficulty of IF and DL. While it is commonly
believed that DL is much harder than IF, we show that the hardness
ratio is roughly a factor of~3 for the current 800-bit range for safe
primes,
much less than previously thought.

\item Last but not least, our computations were performed with the open-source
software \cadonfs\ \cite{cadonfs}. Reproducing our results is possible:
we have set up a companion code repository at
\url{https://gitlab.inria.fr/cado-nfs/records} that
holds the required information to reproduce them.
\end{itemize}

We complement the present work with another record computation,
the factoring of RSA-250, for which we used parameters similar to RSA-240.
Details for this additional record are given at the end of the paper.

This article is organized as follows. We give a brief
introduction to key aspects of NFS in Section~\ref{sec:nfs}.
In Sections~\ref{sec:polyselect} to~\ref{sec:descent+sqrt} we detail the
main steps of NFS, how we chose parameters, and how our computations
proceeded, both for factoring and discrete logarithm.
Section~\ref{sec:simulation} gives further details on the simulation
mechanism that we used in order to predict the running time.
Section~\ref{sec:conclusion} concludes with a comparison with recent
computational records, and a discussion on the relative hardness of the
discrete logarithm and factoring problems.
%

\section{Background on the Number Field Sieve}
\label{sec:nfs}

The Number Field Sieve (NFS) is an algorithmic framework that 
can tackle
either of the two following problems:
\begin{itemize}
    \item Integer factorization (IF): given a composite integer $N$, find a
        non-trivial factorization of $N$.
    \item Discrete logarithm in finite fields (DL): given a prime-order finite
        field\footnote{Variants for non-prime finite fields also exist,
        but are not covered here.} \GF p and a subgroup $G$ of prime
        order $\ell$ within $\GF p^*$, compute a non-trivial
        homomorphism from $G$ to \ZnZ\ell.
        By combining information of this
        kind for various $\ell$, given $g,y\in\GF p^*$, one can compute $x$ such that $g^x=y$.
\end{itemize}
When the need arises, the algorithms for the two problems above are
denoted NFS and NFS-DL, respectively. Most often the acronym NFS is used
for both cases.
Furthermore, in the few cases in this paper where we work with the prime factors of
$N$, we call them $p$ and $q$. Of course, $p$ here shall not be confused
with the prime $p$ of the DLP case. The context allows to avoid the
confusion.

NFS is described in the book~\cite{LeLe93}. So-called ``special''
variants of NFS exist as well, and were historically the first to be developed.
These variants apply when the number $N$ or $p$ has a particular form.
Large computations in these special cases were reported
in~\cite{AC:AFKLO07,EC:FGHT17,AC:KleBosLen14}. In this work, we are concerned only with the
general case (GNFS). The time and space complexity
can be expressed as
$$
L_N(1/3,(64/9)^{1/3})^{1+o(1)}=
\exp\left((64/9)^{1/3}(\log N)^{1/3}(\log\log N)^{2/3}(1+o(1))\right)$$
for factoring.  For discrete logarithms in a subgroup of $\GF p^*$,
$N$ is substituted by $p$ in the above formula.
In both cases, the presence of $(1+o(1))$ in the exponent reveals a
significant lack of accuracy in this complexity estimate, which easily
swallows any speedup or slowdown that would be polynomial in $\log N$.

\begin{figure}
    \begin{center}
{\small
\pgfdeclarelayer{background}
\pgfdeclarelayer{foreground}
\pgfsetlayers{background,main,foreground}

\tikzstyle{sensor}=[draw, fill=blue!20, text width=5em, 
    text centered, minimum height=2.5em]
\tikzstyle{ann} = [above, text width=5em]
\tikzstyle{naveqs} = [sensor, text width=6em, fill=red!20, 
    minimum height=12em, rounded corners]
\tikzstyle{compute}=[draw, fill=blue!20,
    text centered, shape=rectangle]
\tikzstyle{bigcompute}=[draw, fill=blue!20,
    text centered, shape=rectangle,minimum size=4mm]
\tikzstyle{joblabel}=[text centered, text width=5em]
\def\blockdist{2.3}
\def\edgedist{2.5}

    \begin{tikzpicture}[scale=.7,
            every node/.style={transform shape},
            >=stealth]

    \node (p) {$N$};
    \coordinate (gather1)       at ($(p.east)+(.6,0)$);
    \coordinate (ps)            at ($(gather1)+(.6,0)$);
    \coordinate (gather2)       at ($(ps)+(.5,0)$);
    \coordinate (gather3)       at ($(gather2)+(.3,0)$);
    \coordinate (s)             at ($(gather3)+(1,0)$);
    \coordinate (gather4)       at ($(s)+(.9,0)$);
    \coordinate (pos-filtering) at ($(gather4)+(.7,0)$);
    \coordinate (gather5)       at ($(pos-filtering)+(.6,0)$);
    \coordinate (la)            at ($(gather5)+(.8,0)$);
    \coordinate (gather6)       at ($(la)+(.8,0)$);
    \coordinate (pos-sqrt)      at ($(gather6)+(.6,0)$);
    \coordinate (pos-x)         at ($(pos-sqrt)+(1,0)$);

    \draw [-] (p.east) -- (gather1);

    \node[compute] (ps2) at ($(ps)+(0,.33)$){};
    \node[compute] (ps3) at ($(ps)+(0,0)$){};
    \node[compute] (ps4) at ($(ps)+(0,-.33)$){};
    \draw[->] (gather1) -- (ps2.west);
    \draw[->] (gather1) -- (ps3.west);
    \draw[->] (gather1) -- (ps4.west);
    \draw     (ps2.east) -- (gather2);
    \draw     (ps3.east) -- (gather2);
    \draw     (ps4.east) -- (gather2);
    \node (polysel) [above=2.5ex of ps2, joblabel] {polynomial selection};

    \draw [-] (gather2) -- (gather3);

    \node[compute] (s1) at ($(s)+(0,.66)$) {};
    \node[compute] (s2) at ($(s)+(0,.33)$) {};
    \node[compute] (s3) at ($(s)+(0,0)$) {};
    \node[compute] (s4) at ($(s)+(0,-.33)$) {};
    \node[compute] (s5) at ($(s)+(0,-.66)$) {};
    \draw[->] (gather3) -- (s1.west);
    \draw[->] (gather3) -- (s2.west);
    \draw[->] (gather3) -- (s3.west);
    \draw[->] (gather3) -- (s4.west);
    \draw[->] (gather3) -- (s5.west);
    \draw [-] (s1.east) -- (gather4);
    \draw [-] (s2.east) -- (gather4);
    \draw [-] (s3.east) -- (gather4);
    \draw [-] (s4.east) -- (gather4);
    \draw [-] (s5.east) -- (gather4);
    \node[joblabel] at (polysel -| s) {sieving};

    \node[compute] (filtering) at (pos-filtering) {};
    \draw [->] (gather4) -- (filtering.west);
    \draw [-] (filtering) -- (gather5);
    \node[joblabel] at (polysel -| filtering) {filtering};

    \node[bigcompute] (la1) at ($(la)+(0,.5)$) {};
    \node[bigcompute] (la2) at ($(la)+(0,-.5)$) {};
    \draw[->] (gather5) -- (la1.west);
    \draw[->] (gather5) -- (la2.west);
    \draw [-] (la1.east) -- (gather6);
    \draw [-] (la2.east) -- (gather6);
    \node[joblabel] at (polysel -| la) {linear algebra};

    \node[compute] (sqrt) at (pos-sqrt) {};
    \path (polysel -| sqrt) node [joblabel] {square\\ root};
    \draw [->] (gather6.east) -- (sqrt.west);

    \node (x) at (pos-x) {$p, q$};
    \draw [->] (sqrt.east) -- (x.west);

    \begin{pgfonlayer}{background}
        \coordinate (a) at ($(p.west |- polysel.north)+(-.1,.1)$);
        \coordinate (b) at ($(s5.south -| x.east)+(.1,-.3)$);
        \path[fill=yellow!20,rounded corners, draw=black!50] (a) rectangle (b);
    \end{pgfonlayer}

\end{tikzpicture}
}

NFS for factoring: given an RSA modulus $N$, find $p,q$ such that $N=pq$.
\bigskip

        {
\small
\pgfdeclarelayer{background}
\pgfdeclarelayer{foreground}
\pgfsetlayers{background,main,foreground}

\tikzstyle{sensor}=[draw, fill=blue!20, text width=5em, 
    text centered, minimum height=2.5em]
\tikzstyle{ann} = [above, text width=5em]
\tikzstyle{naveqs} = [sensor, text width=6em, fill=red!20, 
    minimum height=12em, rounded corners]
\tikzstyle{compute}=[draw, fill=blue!20,
    text centered, shape=rectangle]
\tikzstyle{bigcompute}=[draw, fill=blue!20,
    text centered, shape=rectangle,minimum size=4mm]
\tikzstyle{joblabel}=[text centered, text width=5em]
\tikzstyle{database}=[
      cylinder,
      cylinder uses custom fill,
      cylinder body fill=yellow!50,
      cylinder end fill=yellow!50,
      shape border rotate=90,
      aspect=.25,
      draw,
      minimum size=1cm
    ]
\def\blockdist{2.3}
\def\edgedist{2.5}

    \begin{tikzpicture}[scale=.7, every node/.style={transform shape},
        >=stealth]

    \node (p) {$p$};
    \coordinate (gather1)       at ($(p.east)+(.6,0)$);
    \coordinate (ps)            at ($(gather1)+(.6,0)$);
    \coordinate (gather2)       at ($(ps)+(.5,0)$);
    \coordinate (gather3)       at ($(gather2)+(.3,0)$);
    \coordinate (s)             at ($(gather3)+(1,0)$);
    \coordinate (gather4)       at ($(s)+(.9,0)$);
    \coordinate (pos-filtering) at ($(gather4)+(.7,0)$);
    \coordinate (gather5)       at ($(pos-filtering)+(.6,0)$);
    \coordinate (la)            at ($(gather5)+(.8,0)$);
    \coordinate (gather6)       at ($(la)+(.8,0)$);
    \coordinate (pos-logdb)     at ($(gather6)+(.6,0)$);

    \draw [-] (p.east) -- (gather1);

    \node[compute] (ps2) at ($(ps)+(0,.33)$){};
    \node[compute] (ps3) at ($(ps)+(0,0)$){};
    \node[compute] (ps4) at ($(ps)+(0,-.33)$){};
    \draw[->] (gather1) -- (ps2.west);
    \draw[->] (gather1) -- (ps3.west);
    \draw[->] (gather1) -- (ps4.west);
    \draw     (ps2.east) -- (gather2);
    \draw     (ps3.east) -- (gather2);
    \draw     (ps4.east) -- (gather2);
    \node (polysel) [above=2.5ex of ps2, joblabel] {polynomial selection};

    \draw [-] (gather2) -- (gather3);

    \node[compute] (s1) at ($(s)+(0,.66)$) {};
    \node[compute] (s2) at ($(s)+(0,.33)$) {};
    \node[compute] (s3) at ($(s)+(0,0)$) {};
    \node[compute] (s4) at ($(s)+(0,-.33)$) {};
    \node[compute] (s5) at ($(s)+(0,-.66)$) {};
    \draw[->] (gather3) -- (s1.west);
    \draw[->] (gather3) -- (s2.west);
    \draw[->] (gather3) -- (s3.west);
    \draw[->] (gather3) -- (s4.west);
    \draw[->] (gather3) -- (s5.west);
    \draw [-] (s1.east) -- (gather4);
    \draw [-] (s2.east) -- (gather4);
    \draw [-] (s3.east) -- (gather4);
    \draw [-] (s4.east) -- (gather4);
    \draw [-] (s5.east) -- (gather4);
    \node[joblabel] at (polysel -| s) {sieving};

    \node[compute] (filtering) at (pos-filtering) {};
    \draw [->] (gather4) -- (filtering.west);
    \draw [-] (filtering) -- (gather5);
    \node[joblabel] at (polysel -| filtering) {filtering};

    \node[bigcompute] (la1) at ($(la)+(0,.5)$) {};
    \node[bigcompute] (la2) at ($(la)+(0,-.5)$) {};
    \draw[->] (gather5) -- (la1.west);
    \draw[->] (gather5) -- (la2.west);
    \draw [-] (la1.east) -- (gather6);
    \draw [-] (la2.east) -- (gather6);
    \node[joblabel] at (polysel -| la) {linear algebra};

    \node[right,database] (logdb) at (pos-logdb) {log db};
    \draw [->] (gather6.east) -- (logdb.west);

    \node(precomputation) at ($(gather4)+(0,-1.25)$) {\textbf{one-off precomputation}};

\coordinate (gather7) at ($(logdb.east)+(1,0)$);
\coordinate (gather8) at ($(gather7)+(.8,0)$);
\coordinate (d) at ($(gather8)+(.6,0)$);
\coordinate (gather9) at ($(d)+(.5,0)$);
\coordinate (pos-x) at ($(gather9)+(1,0)$);

\draw [->] (logdb.east) -- (gather7);
\node[joblabel] (y) at (polysel -| gather7) {$y,g$};
\draw [->] (y) -- (gather7);

\draw [-] (gather7) -- (gather8);

\node[compute] (d2) at ($(d)+(0,.33)$) {};
\node[compute] (d3) at ($(d)+(0,0)$) {};
\node[compute] (d4) at ($(d)+(0,-.33)$) {};
\draw [->] (gather8) -- (d2.west);
\draw [->] (gather8) -- (d3.west);
\draw [->] (gather8) -- (d4.west);
\draw [-] (d2.east) -- (gather9);
\draw [-] (d3.east) -- (gather9);
\draw [-] (d4.east) -- (gather9);
\node[joblabel] at (polysel -| d) {descent};

\node (x) at (pos-x) {$x$};
\draw [->] (gather9) -- (x.west);

    \node at (precomputation -| d)  {\textbf{per-key computation}};

    \begin{pgfonlayer}{background}
        \coordinate (a) at ($(p.west |- polysel.north)+(-.1,.1)$);
        \coordinate (b) at ($(precomputation.south -| logdb.east)+(.2,-.1)$);
        \path[fill=yellow!20,rounded corners, draw=black!50] (a) rectangle (b);
        \coordinate (c) at ($(b.east)+(.1,0)$);
        \coordinate (d) at ($(x.east |- a)+(.1,0)$);
	\path[fill=red!10,rounded corners, draw=black!50] (c) rectangle (d);
    \end{pgfonlayer}

\end{tikzpicture}
}

NFS for DLP: given $g^x \equiv y \bmod p$, find $x$.
\bigskip

\caption{\label{fig:nfs-steps}Main steps of NFS and NFS-DL.}
    \end{center}
\end{figure}

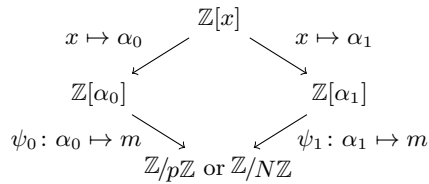
\begin{figure}
\begin{center}
    \begin{tikzpicture}
        \node (t) at (90:1.) {$\bZ [x]$}; 
        \node (l) at (180:1.6) {%
            $\bZ [\alpha_0]$};
            \node (r) at (0:1.6) {$\bZ
            [\alpha_1]$%
            };
            \node (b) at (270:1.) {\ZnZ p or \ZnZ N};
            \draw[->] (t) -- node[above left] {$x\mapsto \alpha_0$\ \ }             (l);
            \draw[->] (t) -- node[above right] {\ $x\mapsto \alpha_1$}       (r);
            \draw[->] (l) -- node[below left,yshift=1ex] {$\psi_0\colon \alpha_0\mapsto m$\ \ }       (b);
            \draw[->] (r) -- node[below right,yshift=1ex] {\ $\psi_1\colon \alpha_1\mapsto m$} (b);
    \end{tikzpicture}
    \caption{\label{fig:nfs-diag}The mathematical setup of NFS.}
\end{center}
\end{figure}

The Number Field Sieve is made up of several independent steps, which are depicted in
Figure~\ref{fig:nfs-steps}.  The first step of NFS, called polynomial
selection, determines a mathematical setup that is well suited to dealing
with the input $N$ (or $p$). That is, we are searching for two irreducible
polynomials $f_0$ and $f_1$ in $\bZ[x]$ that define two algebraic number
fields $K_0=\bQ(\alpha_0)$ and $K_1=\bQ(\alpha_1)$ (with $f_i(\alpha_i)=0$),
subject to some compatibility condition. The resulting maps are depicted in the diagram in
Figure~\ref{fig:nfs-diag}.

To be compatible, $f_0$ and $f_1$ should have a common
root $m$ modulo $p$ (or, likewise, modulo $N$), used in the maps $\psi_0$
and $\psi_1$
in Figure~\ref{fig:nfs-diag}.
This condition is difficult to ensure modulo a composite integer $N$,
and most efficient constructions are limited to choosing $f_0$ as a
linear polynomial, so that $\bZ[\alpha_0]$ is a subring of \bQ. This
leads to commonly used terminology that distinguishes between the
``rational side'' ($f_0$) and the ``algebraic side'' ($f_1$). When
dealing with IF-related considerations, we also use this terminology. In
contrast, for NFS-DL, other constructions exist that take advantage of the
ability to find roots modulo $p$.

Based on the mathematical setup above, the most time-consuming phase of
NFS consists of collecting relations.
We search for elements $a-bx\in\bZ[x]$,
preferably with small coefficients, such that the two integers%
\footnote{A common abuse
of terminology is to use the term ``norm'' to denote $\Res(a-bx,f_i)$,
while in fact the latter coincides with the norm of $a-b\alpha_i$ only
when $f_i$ is monic.}
$\Res(a-bx,f_0)$ and $\Res(a-bx,f_1)$ are \emph{smooth}, i.e., factor
into small prime numbers below some chosen \emph{large prime
bound}.
This
ensures that the ideals $(a-b\alpha_0)\cO_{K_0}$ and
$(a-b\alpha_1)\cO_{K_1}$ both factor into ideals within finite sets
called $\cF_0$ and $\cF_1$.%
\footnote{The terms \emph{smoothness bound} and \emph{factor base} are
fairly standard,
but lead to ambiguous interpretations as we dive into the technical
details of how relation collection is performed. Both are therefore
avoided here,
on purpose.}

The main mathematical obstacle to understanding NFS is that we cannot expect $a-b\alpha_i$
for $i\in\{0,1\}$ to factor into elements of
$\bZ[\alpha_i]$. Only factorization into prime ideals, within the maximal
orders $\cO_{K_i}$, holds. As such,
even given known ideal factorizations of the form
$(a-b\alpha_i)\cO_i=\prod_{\fkp\in\cF_i}\fkp^{e_{\fkp,a,b}}$, it is
impossible to use the diagram in Figure~\ref{fig:nfs-diag} to write a
relation of the following kind (in either \ZnZ N or \ZnZ p, depending on
the context):
\begin{equation}
    \label{eq:mythical-nfs}
\prod_{\fkp\in\cF_0}\psi_0(\fkp^{e_{\fkp,a,b}})
    \mathbin{\ \text{``$\equiv$''}\ }
\prod_{\fkp\in\cF_1}\psi_1(\fkp^{e_{\fkp,a,b}}).
\end{equation}
Indeed the maps $\psi_i$ are defined on elements, not ideals. A prime ideal, or a factorization into prime ideals, does not
uniquely identify an element of $K_i$, because of obstructions related to
the unit group and the class group of the number field $K_i$.  To make an
equality similar to~\eqref{eq:mythical-nfs} work, additional data is
needed, called quadratic characters in the context of
factoring~\cite[§6]{BuLePo93}
and Schirokauer maps in the discrete logarithm
context~\cite{Schirokauer93}. With this additional information,
relations that are functionally equivalent to
Equation~\eqref{eq:mythical-nfs} can be obtained. The general goal of
NFS, once relations have been collected, is:
\begin{itemize}
    \item in the IF context, to combine (multiply) many relations
        together so as to form an equality of squares in \ZnZ N. To achieve
        this goal, it suffices to only keep track of the valuations
        $e_{\fkp,a,b}$ modulo $2$. The right combination can be
        obtained by searching for a left nullspace element of a binary
        matrix;
    \item in the DL context, to view the multiplicative
        relations as additive relations involving unknown logarithms of
        the elements in $\cF_0$ and $\cF_1$, and to solve the corresponding
        linear system for these unknowns. These logarithms, and the
        linear system, are defined
        modulo $\ell$, which is the order of the subgroup of $\GF p^*$
        that we are working in. (If the system were homogeneous, solving it would
        require a right nullspace element of a
        matrix defined modulo $\ell$.)
\end{itemize}

In both cases (factoring and discrete logarithm), we need a linear algebra
calculation. The matrix rows are
vectors of valuations, which are by construction very sparse.  Note
however that record-size computations with NFS typically collect billions of
relations, which is a rather awkward data set to deal with. The linear
algebra step therefore begins with a preprocessing step called
\emph{filtering}, which
carries out some preliminary steps of Gaussian elimination
with the aim of reducing the matrix size significantly while keeping it relatively
sparse.

The linear algebra step that follows is the second most expensive
computational step of NFS. Once the solution to the linear algebra
problem has been found, the final task is to factor $N$ given an equality
of squares modulo $N$, or to compute arbitrary discrete logarithms based
on a database of known logarithms derived from the solution of the linear
system. This final result is obtained via some additional steps which, while
nontrivial, are computationally insignificant compared to the rest of
the computation.

\medskip
NFS can be implemented in software. Our computation was carried out exclusively
with the \cadonfs\ implementation~\cite{cadonfs}.
In
Sections~\ref{sec:polyselect} to~\ref{sec:descent+sqrt}, we examine the
underlying algorithms as well as the
computational specifics of the different steps of NFS.

\section{Polynomial selection}
\label{sec:polyselect}

Polynomial selection can be done using a variety of algorithms, both in
the factoring context~\cite{Kleinjung06,Kleinjung08,Murphy99,BaBoKrZi16} or
in the discrete logarithm context, which allows additional
constructions such as~\cite{JoLe03}. Not all choices of polynomials
$(f_0,f_1)$ that permit the structure in the
diagram in Figure~\ref{fig:nfs-diag} perform equally well. Hence, it is
useful to try many different pairs $(f_0,f_1)$ until a
good one is found. The main optimization criteria are meant to ensure
that on the one hand $\Res(a-bx,f_0)$ and $\Res(a-bx,f_1)$ are somewhat small, 
and on the other hand they are likely to have many small prime factors.

As detailed in~\cite{Kleinjung06}, the most important task in polynomial
selection is to quickly discard the less promising pairs, and efficiently
rank the promising ones with a sequence of estimators, from coarse-grained
estimators to finer-grained ones. Finally, a small-scale
sieving test can be run in order to select the polynomial pair to use
for the rest of the computation among the final set of candidates.

We followed exactly this approach, using the Murphy-$E$
value from \cite{Murphy99}
as well as
the modified estimator $E'$ suggested in \cite{david:hal-02151093} as estimators.
We performed sample sieving for a few dozen of the best candidates.
\cadonfs\ can do this easily with
the \verb|random-sample| option.

\subsection{Computational data} 

\subsubsection{RSA-240.}

We
used Kleinjung's algorithm \cite{Kleinjung06,Kleinjung08}, with
improvements from \cite{BaBoKrZi16}, and
searched for a polynomial pair with $\deg f_0=1$ and $\deg f_1=6$.
We forced the leading coefficient $f_{1,6}$ of $f_1$ to be divisible
by $110880$, to ensure higher divisibility by $2, 3, 5, 7, 11$.
The parameter $P$ used (see \cite{Kleinjung08}) was $P = 2\cdot 10^7$
and we searched up to $f_{1,6} = 2 \cdot 10^{12}$.


The cost of the search was about 76 core-years. It was distributed over
many computer clusters, and took only 12 days of calendar time.
We kept 40 polynomial pairs:
the top 20 with the best Murphy-$E$ values, and the
top 20 with the modified $E'$ value~\cite{david:hal-02151093}.
After some sample sieving, two clear winners emerged, and distinguishing between both was actually
quite hard. 
In the end we chose the
one optimized for the classical Murphy-$E$ value,
with $\left|\Res(f_0,f_1)\right| = 120 N$:
{\small
\begin{eqnarray*}
  f_1 &=& {10853204947200}\,x^6
    - {221175588842299117590564542609977016567191860}\,
    \\
    &-& {4763683724115259920}\,x^5
    + {1595712553369335430496125795083146688523}\,x
    \\
    &-& {6381744461279867941961670}\,x^4
    + {179200573533665721310210640738061170}\,x^2
    \\
    &+& {974448934853864807690675067037}\,x^3 \\
  f_0 &=& {17780390513045005995253}\,x 
    - {105487753732969860223795041295860517380}\,
\end{eqnarray*}
}
For each polynomial $f_i(x)$, we denote by $F_i(x,y)=y^{\deg f_i}f_i(x/y)$ the corresponding
homogeneous bivariate polynomial.

\subsubsection{DLP-240.}

As in \cite{EC:KDLPS17}, we used the Joux-Lercier selection
algorithm~\cite{JoLe03}, searching for a pair $(f_0,f_1)$ with
$f_1$ of degree $d$ with small coefficients, and $f_0$ of
degree $d-1$ with coefficients of order $p^{1/d}$. As in~\cite{EC:KDLPS17},
we used $d=4$ which is optimal for this size, with coefficients of
$f_1$ bounded by 150 in absolute value, compared to 165 in~\cite{EC:KDLPS17}.

The cost of the search was about 152 core-years,
and only about 18 days of calendar time.
We kept the 100 best pairs according to their Murphy-$E$ value, and
chose the winning polynomial pair based on the results of sample
sieving.
As in the RSA-240 case, the very best handful of polynomials provided
almost identical yields. 
We
ended up using the following
pair,
with $\left|\Res(f_0,f_1)\right| = 540 p$:
{\small
\begin{eqnarray*}
  f_1 &=& 39x^4+126x^3+x^2+62x+120 \\
  f_0 &=& 286512172700675411986966846394359924874576536408786368056 x^3 \\
    &+& 24908820300715766136475115982439735516581888603817255539890 x^2 \\
    &-& 18763697560013016564403953928327121035580409459944854652737 x \\
    &-& 236610408827000256250190838220824122997878994595785432202599
\end{eqnarray*}
}
%
Note that although there is a clear correlation between the efficiency of
a polynomial pair and its Murphy-$E$ value, the ranking is definitely not
perfect \cite{david:hal-02151093}; in particular, the top scoring polynomial
pair according to Murphy-$E$ finds 10\% fewer relations than the above one.


\section{Relation collection}
\label{sec:sieving}

The relation collection uses a technique called \emph{lattice
sieving}~\cite{Pollard93b}.
Lattice sieving borrows
from the terminology of \emph{special-$q$
sieving}~\cite{C:DavHol83}. 
We call special-\fkq\ ideals a
large set of ideals of one of the two number fields\footnote{It is
possible to mix special-\fkq\ ideals from both number fields, as done
in~\cite{EC:FGHT17}, or even hybrid special-\fkq\ involving
contributions from both sides.}.  For each such special-\fkq, the search
for relations is done among the pairs $(a,b)$ such that the prime ideals
dividing \fkq\ appear in the factorization\footnote{By
factorization, we implicitly mean ``numerator of the factorization''.
Furthermore we factor ideals such as $(a-b\alpha)\cO_{K}$, yet
the maximal order $\cO_{K}$ is generally too expensive to compute.
It turns out that if $\Res(a-bx,f)$ is smooth and fully
factored, then it is easy to do. How to deal with these technicalities is well known, and not
discussed here (see~\cite[chapters~4 and~6]{Cohen93}).}
of $(a-b\alpha_0)\cO_{K_0}$
(or $(a-b\alpha_1)\cO_{K_1}$). These $(a,b)$ pairs form a lattice
$\cL_\fkq$ in $\bZ^2$, which depends on \fkq. Let $(\vec u,\vec v)$
be a Gauss-reduced basis of $\cL_\fkq$.  To enumerate small points in
$\cL_\fkq$, we consider small linear combinations of the form $(a,b)=i\vec
u+j\vec v$.
In order to search for good pairs $(a,b)$ in $\cL_q$, we use the change
of basis given by $(\vec u,\vec v)$, and instead search for good pairs $(i,j)$
such that both $\Res(a-bx,f_0)$ and $\Res(a-bx,f_1)$ are
smooth.

\begin{figure}
    \subfigbottomskip 0pt
\def\do[#1](#2),(#3){
    \coordinate (u0) at (#2);
    \coordinate (u1) at (#3);
    \coordinate (u01a) at ($(u1)+(u0)$);
    \coordinate (u0m) at ($(0,0)-(u0)$);
    \coordinate (u01b) at ($(u1)-(u0)$);
    \fill[#1,opacity=.25] (0,0) -- (u0) -- (u01a) -- (u01b) -- (u0m) -- cycle;
    \draw (0,0) -- (u0) -- (u01a) -- (u01b) -- (u0m) -- cycle;
}
\begin{center}
    \begin{minipage}[b]{.33\textwidth}
\begin{center}
    \subfigure[][]{\label{fig:lattice-sieving-rectangles-a}%
\begin{tikzpicture}[x=4.75ex,y=4.75ex,every node/.style={only marks},baseline=0]
    \fill[use as bounding box,pattern=dots] (-3,-2) rectangle (3,2);
    \begin{scope}[xshift=-.15ex]
    \node[minimum size=3pt,inner sep=0pt,black,fill=red,circle] at (0,0) {};
        \do[red](-0.75, 0.19),(-0.016, 1.5)
        \do[yellow](-0.50, -0.25),(-0.75, 1.5)
        \do[green](0.094, -0.75),(-1.0, -0.19)
        \do[blue](0.38, 0.38),(1.5, -1.5)
    \end{scope}
\end{tikzpicture}
}\\[0pt]
\end{center}
\end{minipage}
\hfil
\hfil
\begin{minipage}[b]{.25\textwidth}
\begin{center}
    \subfigure[][]{\label{fig:lattice-sieving-rectangles-b}%
    \def\row#1;{\path plot[boldsieve] coordinates{#1};}%
\begin{tikzpicture}[
        x=.75em,
        y=4ex,
        scale=0.5,
        baseline=0,
        thick,
        sieve/.style={only marks,mark=*,mark options={color=darkgreen}},
        boldsieve/.style={sieve, mark size=4pt},
        blueboldsieve/.style={boldsieve, mark options={color=darkblue},mark size=4pt},
    ]
    \path[use as bounding box] (-10,0) rectangle (10,10);
\colorlet{darkgreen}{green!50!black}
\colorlet{darkblue}{blue!50!black}
\path plot[sieve] coordinates{
(-8.6, 0.0)
(-4.3, 0.0)
(0.0, 0.0)
(4.3, 0.0)
(8.6, 0.0)
(-9.9, 1.0)
(-5.6, 1.0)
(-1.3, 1.0)
(3.0, 1.0)
(7.3, 1.0)
(-6.9, 2.0)
(-2.6, 2.0)
(1.7, 2.0)
(6.0, 2.0)
(-8.2, 3.0)
(-3.9, 3.0)
(0.4, 3.0)
(4.7, 3.0)
(9.0, 3.0)
(-9.5, 4.0)
(-5.2, 4.0)
(-0.9, 4.0)
(3.4, 4.0)
(7.7, 4.0)
(-6.5, 5.0)
(-2.2, 5.0)
(2.1, 5.0)
(6.4, 5.0)
(-7.8, 6.0)
(-3.5, 6.0)
(0.8, 6.0)
(5.1, 6.0)
(9.4, 6.0)
(-9.1, 7.0)
(-4.8, 7.0)
(-0.5, 7.0)
(3.8, 7.0)
(8.1, 7.0)
(-6.1, 8.0)
(-1.8, 8.0)
(2.5, 8.0)
(6.8, 8.0)
(-7.4, 9.0)
(-3.1, 9.0)
(1.2, 9.0)
(5.5, 9.0)
(9.8, 9.0)
};
\fill[yellow,opacity=0.5](-7,0) -- (-7,8) -- (7,8) -- (7,0);
\draw(-7,0) -- (-7,8) -- (7,8) -- (7,0);
    \row (-4.3, 0.0) (0.0, 0.0) (4.3, 0.0);
\row (-1.3,1.0) (-5.6,1.0) (-1.3, 1.0) (3.0, 1.0);
\row (-2.6,2.0) (-6.9,2.0) (-2.6, 2.0) (1.7, 2.0) (6.0, 2.0);
\row (-3.9,3.0) (0.4, 3.0) (4.7, 3.0);
\row(-5.2, 4.0) (-0.9, 4.0) (3.4, 4.0);
\row(-6.5, 5.0) (-2.2, 5.0) (2.1, 5.0) (6.4, 5.0);
\row(-3.5, 6.0) (0.8, 6.0) (5.1, 6.0);
\row(-4.8, 7.0) (-0.5, 7.0) (3.8, 7.0);
\end{tikzpicture}
}\\[0pt]
\end{center}
\end{minipage}
\hfil
\begin{minipage}[b]{.25\textwidth}
\begin{center}
    \subfigure[][]{\label{fig:lattice-sieving-rectangles-c}%
\def\row#1;{\path plot[boldsieve] coordinates{#1};}%
    \begin{tikzpicture}[
        x=.75em,
        y=4ex,
        scale=0.5,
        baseline=0,
        thick,
        sieve/.style={only marks,mark=*,mark options={color=darkgreen}},
        boldsieve/.style={sieve, mark size=4pt},
        blueboldsieve/.style={boldsieve, mark options={color=darkblue},mark size=4pt},
    ]
        \path[use as bounding box] (-10,0) rectangle (10,10);
\colorlet{darkgreen}{green!50!black}
\colorlet{darkblue}{blue!50!black}
\path plot[sieve] coordinates{
(0.0, 0.0)
(3.0, 1.0)
(6.0, 2.0)
(-8.2, 3.0)
(9.0, 3.0)
(-5.2, 4.0)
(-2.2, 5.0)
(0.8, 6.0)
(3.8, 7.0)
(6.8, 8.0)
(-7.4, 9.0)
(9.8, 9.0)
};
\fill[yellow,opacity=0.5](-7,0) -- (-7,8) -- (7,8) -- (7,0);
\draw(-7,0) -- (-7,8) -- (7,8) -- (7,0);
\row(0.0, 0.0);
\row(3.0, 1.0);
\row(6.0, 2.0);
\row(-5.2, 4.0);
\row(-2.2, 5.0);
\row(0.8, 6.0);
\row(3.8, 7.0);
\end{tikzpicture}
}\\[0pt]
\end{center}
\end{minipage}
\end{center}
\vskip-1.5\baselineskip
\caption{\label{fig:lattice-sieving-rectangles}\subref{fig:lattice-sieving-rectangles-a}: Examples of $(i,j)$
rectangles for various lattices $\cL_\fkq$ within the $(a,b)$ plane.
    \subref{fig:lattice-sieving-rectangles-b}: Sieving for a prime
    $p$ in the $(i,j)$ rectangle.
    \subref{fig:lattice-sieving-rectangles-c}: Sieved
    locations can be quite far apart, and accessing them naively can
    incur a significant memory access penalty.}
\end{figure}
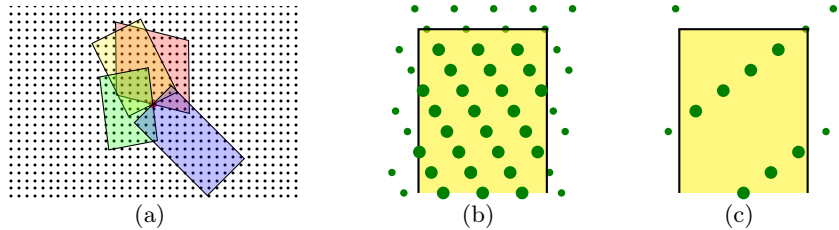

The set of explored pairs $(i,j)$ is called the \emph{sieve area}, which
we commonly denote by \cA. 
For performance it is best to
have \cA\ of the form $[-I/2,I/2)\times[0,J)$ for some integers
$I$ and $J$, and $I$ a power of two. This implies that as we consider multiple special-\fkq{}s,
the sieved rectangles drawn in
Figure~\ref{fig:lattice-sieving-rectangles-a} (whose intersections with
$\bZ^2$ most often have very few common points,
since divisibility conditions are distinct)
lead us to implicitly consider $(a,b)$ pairs that
generally have small norm,
but are not constrained to some area that has been defined a
priori.  In fact, various strategies can be used to
make small adjustments to the sieve area depending on \fkq~in order to limit
the spread of the zones reached in
Figure~\ref{fig:lattice-sieving-rectangles-a}.

Relation collection finds pairs $(a,b)$ (or, equivalently,
pairs $(i,j)$) such that two smoothness conditions hold simultaneously
(see §\ref{sec:nfs}).
We thus have two sides to consider. In the description below, we use \cF\
to denote either $\cF_0$ or $\cF_1$, as the same processing can be
applied to both sides. Likewise, we use $f$ and $\alpha$ to denote either
$f_0$ and $\alpha_0$, or
$f_1$ and $\alpha_1$.

One of the efficient ways to find the good pairs $(i,j)$ is to use a
\emph{sieving} procedure. 
Let \fkp\ be a moderate-size prime ideal in \cF, subject to limits on
$\left|\Norm\fkp\right|$ that will be detailed later.
Identify the locations $(i,j)$ such that
$\fkp\mid(a-b\alpha)$. These locations again form a lattice in the
$(i,j)$ coordinate space, as seen in
Figures~\ref{fig:lattice-sieving-rectangles-b}--\subref{fig:lattice-sieving-rectangles-c} and hence
implicitly a sub-lattice of $\cL_\fkq$.  Record the corresponding
contribution in an array cell indexed by $(i,j)$. Repeat this process for many (not
necessarily all) prime
ideals in $\cF$, and keep the array cells whose
recorded contribution is closest to the value
$\left|\Res(a-bx,f)\right|$: those are the most
promising, i.e., the most likely to
be smooth on the side being sieved. Proceed similarly for the other side, and check the few
remaining $(a,b)$ pairs for smoothness. Note that as \fkp\ varies, the
set of locations where \fkp\ divides $(a-b\alpha)$ becomes sparser (see
Figure~\ref{fig:lattice-sieving-rectangles-c}), and
dedicated techniques must be used to avoid large memory access
penalties.

An alternative approach is to identify the good
pairs $(i,j)$ with product trees, using ``batch smoothness detection'' as explained in~\cite{djb-sf-2002}.
Among a
given set of norms,
determine their smooth part by computing their $\gcd$ with the
product of the norms of all elements of $\cF$ at once. This is
efficient because it can be done while taking advantage of asymptotically
fast algorithms for multiplying integers.  This
approach was used profitably for the previous 768-bit DLP
record~\cite{EC:KDLPS17}.

Among the mind-boggling number of parameters that influence the sieving
procedure, the most important choices are the
following.
\begin{itemize}
    \item The large prime bound that determines the set $\cF$.
        These
        bounds (one on each side) define the
        ``quality'' of the relations we are looking for.
        \cadonfs\ uses the notation $\texttt{lpb}$ for these
        bounds.
    \item The \fkq-range and the size of the sieve area $\#\cA$. This
        controls how many special-\fkq{}s we consider, and how much work
        is done for each. The amount of work can also vary depending on
        the norm of \fkq. The ratio between the norms of the smallest
        and largest special-\fkq\ is important to examine: a large
        ratio increases the likelihood that the same relations are obtained from
        several different special-\fkq\ (called \emph{duplicates}),
        and causes diminishing returns.
        Enlarging the sieve area increases the yield per special-\fkq,
        but also with diminishing returns for the larger area, and costs extra memory. In
        order to collect the expected number of relations, it is
        necessary to tune these parameters.
    \item The size of the prime ideals $\fkp\in\cF$ being sieved for, 
        and more generally \emph{how} (and
        \emph{if}) these ideals are sieved. \cadonfs\ uses the notation $\texttt{lim}$ for this
        upper bound, and we refer to it as the \emph{sieving (upper)
        bound}.
        As a rule of thumb, we should sieve with prime
        ideals that are no larger than the size of the sieve area, 
        so that
        \emph{sieving} actually makes sense. The inner details of the
        lattice sieving implementation also define how sieving is
        performed, e.g., how we transition between algorithms in
        different situations like those depicted in
        Figures~\ref{fig:lattice-sieving-rectangles-b}
        and~\ref{fig:lattice-sieving-rectangles-c}, along with more
        subtle distinctions. This has a significant impact on the amount of
        memory that is necessary for sieving.\\
        When sieving is replaced by batch
        smoothness detection, we also use the notation $\texttt{lim}$ to
        denote the maximum size of primes that are detected with product
        trees.
    \item Which criteria are used to decide that $(a,b)$ are
        ``promising'' after sieving, and the further processing that is
        applied to them.  Typically, sieving identifies a smooth part of
        $\Res(a-b\alpha,f)$, and a remaining unfactored part
        (cofactor).
        Based on the
        cofactor size, one must decide whether it makes sense to seek its
        complete factorization into elements of \cF.
        In this case \cadonfs\ uses the Bouvier-Imbert mixed
        representation \cite{PKC:BouImb20}.
        Any prime ideal that appears in this ``cofactorization'' is called
        a large prime. By construction, large primes are between the sieving bound
        and the large prime bound.
\end{itemize}
\subsection{Details of our relation search}

One of the key new techniques we adopt in our experiments is how we organize the
relation search. The picture is quite different in the two cases. For IF, 
this phase is the most costly, and can therefore be
optimized more or less independently of the others. On the other hand
for DL, the linear algebra becomes the bottleneck by a large margin if the
parameters of the relation search are chosen without considering the
size of the matrix they produce.

The first component that we adjust is the family of special-{\fkq}s that we
consider. In the DL case, a good strategy to help the filtering and have
a smaller matrix is to try to limit the number of
large ideals involved in each relation as much as possible. The approach taken
in~\cite{EC:KDLPS17} was to have special-{\fkq}s\ that stay small, and therefore
to increase the sieve area $\#\cA$, which comes at a cost. We instead chose
 to use \emph{composite special-{\fkq}s} (special-\fkq\
ideals with composite norm), reviving an idea that was originally proposed
in~\cite{Kleinjung06b} in the factoring case
to give estimates for factoring RSA-1024. This idea was extended in~\cite[Section 4.4]{cryptoeprint:2017:758} to the discrete
logarithm case, but to our knowledge it was never used in any practical
computation.
We
pick composite special-{\fkq}s that are larger than the large prime bound, but whose
prime factors are small, and do not negatively influence the
filtering. Because there are many of them, this no longer requires a large sieve area,
so that we can tune it according to other tradeoffs.

In the context of IF, we chose the special-{\fkq}s more classically, some
of them below \texttt{lim}, and some of them between
\texttt{lim} and \texttt{lpb}.
\medskip

\begin{figure}
\textbf{RSA-240\string:}

\noindent
\begin{tikzpicture}
    \coordinate (zz) at (0,0);
    \coordinate (ztop) at (0,.3cm);
    \coordinate (zlab) at (0,.4cm);
    \coordinate (zsub) at (0,-0.15);
    \coordinate (zdeep) at (0,-5.75\baselineskip);
    \coordinate (lim1) at (.36\textwidth,0);
    \coordinate (q0) at (.2\textwidth,0);
    \coordinate (q1) at (.6\textwidth,0);
    \coordinate (lpb) at (.82\textwidth,0);
    \coordinate (far right) at (\textwidth,0);
    \fill[\ablue]   (lim1) rectangle (q0 |- ztop) -- (lim1);
    \fill[\aorange] (lim1) rectangle (q1 |- ztop) -- (lim1);
    \node[left] (fkq) at (far right) {\fkq};
    \node[left] at (\textwidth,-\baselineskip) {(prime)};
    \coordinate (qqq) at (fkq.west);
    \draw[->] (0,0) -- (qqq);
    \draw[-] (ztop) -- (lpb |- ztop);
    \draw (0,0) -- (zlab);
    \node[anchor=south] at (q0 |- zlab) {$2^{29.6}$};
    \draw (lim1 |- ztop) -- (lim1 |- zsub);
    \node[anchor=south] at (lim1 |- zlab) {$2^{31}$};
    \node[anchor=north] (lim1 text) at (lim1 |- zsub) {$\texttt{lim}_1$};
    \node[anchor=south] at (q1 |- zlab) {$2^{32.8}$};
    \draw (lpb) -- (lpb |- ztop);
    \node[anchor=south] (lpb labels) at (lpb |- zlab) {$2^{36}\smash,2^{37}$};
    \path[use as bounding box](zdeep) rectangle (lpb labels.north -| far right);
    \node[anchor=north] at (lpb |- zsub) {$\texttt{lpb}_{0,1}$};
    \draw (lpb) -- ($(lpb)+(0,-0.1cm)$);
    \coordinate (under lim1) at ($(lim1 text.south)+(0,-1ex)$);
    \draw[decorate,decoration={brace,amplitude=0.3cm},thick] (lim1) -- (q0);
    \coordinate(uleft) at ($(under lim1)+(-.75em,0)$);
    \coordinate(uright) at ($(under lim1)+(.75em,0)$);
    \node[fit=(zdeep)(uleft)] (Ltext)
    {\raggedright
              $\left|\Norm\fkq\right| < \texttt{lim}_1$: we allow\\
              \quad 2 large primes on side 0,\\
              \quad 3 large primes on side 1.\\
              We sieve on both sides.\\
     };
    \draw[decorate,decoration={brace,amplitude=0.3cm},thick] (q1) -- (lim1);
    \node[fit=(uright)(zdeep -| far right)] (Rtext) {%
        \raggedright
        $\left|\Norm\fkq\right| > \texttt{lim}_1$:\\
        we allow
        2 large primes on each side.\\
        ($\fkq$ counts as an additional large prime on side 1.)\\
        We sieve on side 1 and batch on side 0.\\};
    \draw (under lim1) -- (lim1 |- zdeep);
    \node[anchor=south] at (zlab) {\rlap {$2^0$}};
\end{tikzpicture}
\smallskip

    \textbf{DLP-240:}

\noindent
\begin{tikzpicture}
    \coordinate (zz) at (0,0);
    \coordinate (ztop) at (0,.3cm);
    \coordinate (zlab) at (0,.4cm);
    \coordinate (zsub) at (0,-0.15);
    \coordinate (zdeep) at (0,-5\baselineskip);
    \coordinate (qmin) at (.06\textwidth,0);
    \coordinate (qmax) at (.2\textwidth,0);
    \coordinate (lim1) at (.3\textwidth,0);
    \coordinate (q0) at (.55\textwidth,0);
    \coordinate (q1) at (.82\textwidth,0);
    \coordinate (lpb) at (.45\textwidth,0);
    \coordinate (far right) at (\textwidth,0);
    \fill[gray]  (qmin|-ztop) rectangle (qmax);
    \fill[\ared] (q0|-ztop) rectangle (q1);
    \node[left] (fkq) at (far right) {\fkq};
    \node[left] at (\textwidth,-\baselineskip) {(composite)};
    \coordinate (qqq) at (fkq.west);
    \draw[->] (0,0) -- (qqq);
    \draw[-] (ztop) -- (ztop -| q1);
    \draw (zz) -- (zlab);
\node[anchor=south] at (qmin |- zlab) {$2^{13}$};
\node[anchor=south] at (qmax |- zlab) {$2^{26.5}$};
    \draw (lim1 |- ztop) -- (lim1 |- zsub);
\node[anchor=south] at (lim1|-zlab) {$2^{29}\smash,2^{28}$};
    \node[anchor=north] at (lim1|-zsub) {\strut$\texttt{lim}_{0,1}$};
    \draw (lpb|-ztop) -- (lpb|-zsub);
    \node[anchor=south] (lpb labels) at (lpb|-zlab) {$2^{35}$};
    \path[use as bounding box](zdeep) rectangle (lpb labels.north -| far right);
\node[anchor=north] at (lpb|-zsub) {\strut$\texttt{lpb}_{0,1}$};
\node[anchor=south] at (q0|-zlab) {$2^{37.1}$};
\node[anchor=south] at (q1|-zlab) {$2^{38.1}$};
    \coordinate (qmiddle) at ($(qmin)!.5!(qmax)$);
    \coordinate (q01) at ($(q0)!.5!(q1)$);
    \node[anchor=north,text centered,text width=12em] (primefact) at
    (qmiddle |- zsub) {\strut$\mathfrak{q}_i,\mathfrak{q}_j$\\
(prime factors of $\fkq$)};
    \draw[decorate,decoration={brace,amplitude=0.3cm},thick] (qmax) --
    node[anchor=north,below=0.4cm] {} (qmin);
    \draw[decorate,decoration={brace,amplitude=0.3cm},thick] (q1) --
    node[anchor=north,below=0.4cm] {} (q0);
    \node[anchor=north,text centered] at
    (q01 |- zsub) {\strut$\fkq=\fkq_i\fkq_j$};
    \node[anchor=south east] at (far right |- zdeep) {
    \begin{tabular}{l}
    We allow 2 large primes on each side.\\
    (The factors of $\fkq$ are not large primes.)\\
    We sieve on side 0 and batch on side 1.
    \end{tabular}};
\node[anchor=south] at (zlab) {\rlap{$2^0$}};
\end{tikzpicture}
\caption{Position of special-\fkq{} ranges with respect to the sieving
bounds \texttt{lim} and the large prime bounds \texttt{lpb}
(values not to scale).
For RSA-240, there are 2 distinct sub-ranges with different kinds of
relations that are sought, with different strategies. For DLP-240, the
special-\fkq\ range is well beyond the large prime bound, thanks to the use of
composite special-\fkq{}s.
}
\label{fig:spq}
\end{figure}
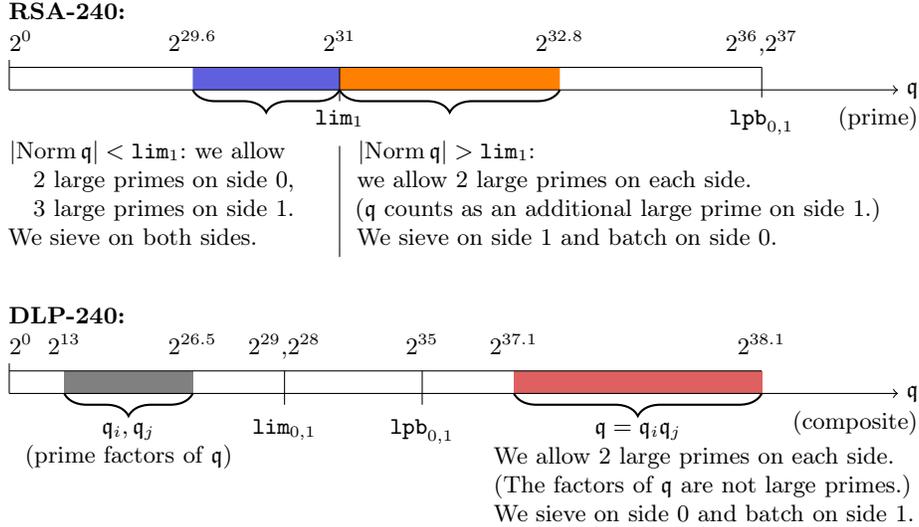

Another important idea is to adapt the relation search strategy depending
on the type of special-\fkq\ we are dealing with and on the quality of the
relations that are sought. A graphical representation of these strategies
is given in Figure~\ref{fig:spq}.

In the DL case, we want to limit as much as possible the
size of the matrix and the cost of the linear algebra.
To this end, we used small sieving bounds, and
allowed only two large primes on each side, with rather small
large prime bounds.
These choices have additional advantages:
a very small number of $(i,j)$ candidates survive the sieving on
the $f_0$-side (the side leading to the largest norms), so that
following~\cite{EC:KDLPS17}, we
skipped sieving on the $f_1$-side entirely and used factorization
trees to handle the survivors, thus saving time and memory by about a
factor of~2 compared to sieving on both sides.

In the IF case, the same idea can also be used, at least to some extent. The
first option would be to have large prime bounds and allowed number of
large primes that follow the trend of
previous factorization records. Then the number of survivors of
sieving on one side is so large that it is not possible to use
factorization trees on the other side, and we have to sieve on both sides.
The other option is to reduce the number of large primes on
the algebraic side (the more difficult side), so that after sieving on this side
there are fewer survivors and we can use factorization trees. Of
course, the number of relations per special-\fkq\ will be reduced, but on
the other hand the cost of finding them is reduced by about a factor of~2.
In our RSA-240 computation, we found that neither option appeared to be definitively optimal, 
and after numerous simulations, we chose
to apply the traditional strategy for the small {\fkq}s (below the
sieving bound \texttt{lim}) and the new strategy for the larger ones.

\subsection{Distribution and parallelization}
\label{subsec:dist-par}

For large computations such as the ones reported in this article, the
relation collection step is a formidable computing effort, and it is also
embarrassingly parallel: as computations for different special-{\fkq}s are
independent, a large number of jobs can run simultaneously, and need no
inter-process communication.  A large amount of computing power must be
harnessed in order to complete this task in a reasonable amount of time.
To this end, several aspects are crucially important.

\subsubsection{Distribution.}

First, the \emph{distribution} of the work is seemingly straightforward.
We may divide the interval $[q_{\text{min}},q_{\text{max}})$ into
sub-intervals of any size we see fit, and have independent jobs process
special-{\fkq}s whose norm lie in these sub-intervals. This approach,
however, needs to be refined if we use multiple computing facilities with
inconsistent software (no common job scheduler, for instance),
inconsistent hardware, and intermittent availability, possibly resulting
in jobs frequently failing to complete. On several of the computing
platforms we had access to, we used so-called \emph{best-effort} jobs, that
can be killed anytime by other users' jobs. This approach means that it
is necessary to keep track of all ``work units'' that have been assigned
at any given point in time, and reliably collect all results from clients. For the
computations reported in this article, the machinery implemented in \cadonfs\ was sufficient. It
consists of a standalone server where each work unit follows a state
machine with the following states: AVAILABLE (a fresh work unit submitted
for processing), ASSIGNED (when a client has
asked for work), OK (result successfully uploaded to server), ERROR
(result failed a sanity check on server, or client error), and CANCELED
(work unit timed out, presumably because the client went offline). Work
units that reach states ERROR or CANCELED are resubmitted up to a few
times, to guard against potential infinite loops caused by software
bugs. This approach was sufficient to deal with most
computation mishaps, and the few remaining ``holes'' were filled manually.

\subsubsection{Parallelization.}
\label{subsubsec:parallelization}

The second crucial aspect is parallelism. The lattice sieving algorithm
that we use in relation collection is not, in itself, easy to
parallelize. Furthermore, it is a memory-intensive computation that is quite
demanding in terms of both required memory and  memory throughput.  In
the most extreme case, having many CPU cores is of little help if the
memory throughput is the limiting factor. The following
(simplified)
strategy was used to run the lattice sieving program at the whole machine
level.
\begin{itemize}
    \item Given the program parameters, determine the amount of memory
        $m$ that is needed to process one special-\fkq.
%
        On a machine with $v$ virtual cores and memory $M$, determine the maximal
        number $s$ of sub-processes and the number of threads $t$ per
        sub-process such that $sm\leq M$, $st=v$, and
        $t$ is a meaningful subdivision of the machine.
        This strategy of maximizing
        $s$ is meant to take advantage of coarse-grained
        parallelism.
    \item Each of the $s$ sub-processes is bound to a given set of $t$
        (virtual) cores of the machine, and handles one special-\fkq\ at
        a time.
\end{itemize}

For each special-\fkq, sub-processes function as follows.
First divide the set of $\fkp\in\cF$ for which we sieve (that is, whose norm is
less than \texttt{lim})
into many \emph{slices} based on several criteria (bounded
slice size, constant value for $\lfloor\log\left|\Norm\fkp\right|\rceil$, same number of
conjugate ideals of $\fkp$). The largest sieved prime ideals in $\cF_0$ have somewhat
rare hits (as in Figure~\ref{fig:lattice-sieving-rectangles-c}).  We
handle them with so-called ``bucket sieving'', which proceeds in two
phases that are parallelized differently:
\begin{itemize}
    \item ``fill buckets'': slices are processed in parallel, and
        ``updates'' are precomputed and appended to several lists, one
        for each ``region'' of the sieve area. These lists are called
        ``buckets''. A region is typically 64kB in size. In order to
        avoid costly concurrent writes, several independent sets of
        buckets can be written to by threads working on different slices.
    \item ``apply buckets'': regions are processed in parallel.  This
        entails reading the information from ``fill buckets'', that is,
        the updates stored in the different lists. Together with this
        second phase of the computation, we do everything that is easy to
        do at the region level: sieve small prime ideals, compute
        $\log\left|\Res(a-bx,f_0)\right|$, and determine whether the
        remaining cofactor is worth further examination.
\end{itemize}

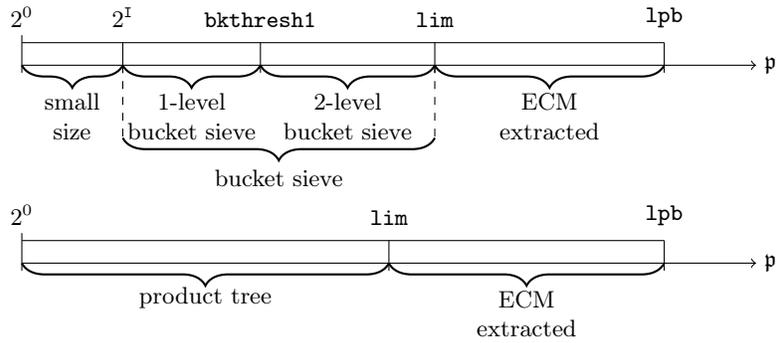
\begin{figure}
  \centering
  \begin{tikzpicture}
  \draw[->] (0,0) -- (0.8\textwidth,0);
  \node[anchor=west] at (0.8\textwidth,0) {$\fkp$};
\draw[-] (0,0.3cm) -- (0.7\textwidth,0.3cm);
\draw (0.0cm,-0.1cm) -- (0.0cm,0.4cm) ;
\node[anchor=south] at (0.0cm,\baselineskip) {$2^0$};
\draw (0.7\textwidth,-0.1cm) -- (0.7\textwidth,0.3cm) ;
\node[anchor=south] at (0.7\textwidth,\baselineskip) {\texttt{lpb}};
\draw (0.11\textwidth,-0.1cm) -- (0.11\textwidth,0.3cm) ;
\node[anchor=south] at (0.11\textwidth,\baselineskip) {$2^{\texttt{I}}$};
\draw (0.26\textwidth,-0.1cm) -- (0.26\textwidth,0.3cm) ;
\node[anchor=south] at (0.26\textwidth,\baselineskip) {\texttt{bkthresh1}};
\draw (0.45\textwidth,-0.1cm) -- (0.45\textwidth,0.3cm) ;
\node[anchor=south] at (0.45\textwidth,\baselineskip) {\texttt{lim}};

\draw[decorate,decoration={brace,mirror,amplitude=0.3cm},thick]
(0,0) --
node[anchor=north,below=0.2cm]{\begin{tabular}{c} small \\ size \end{tabular}}
(0.11\textwidth,0);
\draw[decorate,decoration={brace,mirror,amplitude=0.3cm},thick]
(0.11\textwidth,0) --
node[anchor=north,below=0.2cm]{\begin{tabular}{c} 1-level \\ bucket sieve \end{tabular}}
(0.26\textwidth,0);
\draw[decorate,decoration={brace,mirror,amplitude=0.3cm},thick]
(0.26\textwidth,0) --
node[anchor=north,below=0.2cm]{\begin{tabular}{c} 2-level \\ bucket sieve \end{tabular}}
(0.45\textwidth,0);
\draw[decorate,decoration={brace,mirror,amplitude=0.3cm},thick]
(0.11\textwidth,-2.5\baselineskip) --
node[anchor=north,below=0.3cm]{bucket sieve}
(0.45\textwidth,-2.5\baselineskip);
\draw[dashed] (0.11\textwidth,0) -- (0.11\textwidth,-2.5\baselineskip);
\draw[dashed] (0.45\textwidth,0) -- (0.45\textwidth,-2.5\baselineskip);
\draw[decorate,decoration={brace,mirror,amplitude=0.3cm},thick]
(0.45\textwidth,0) --
node[anchor=north,below=0.2cm]{\begin{tabular}{c} ECM \\ extracted \end{tabular}}
(0.7\textwidth,0);
\end{tikzpicture}

\begin{tikzpicture}
  \draw[->] (0,0) -- (0.8\textwidth,0);
  \node[anchor=west] at (0.8\textwidth,0) {$\fkp$};
\draw[-] (0,0.3cm) -- (0.7\textwidth,0.3cm);
\draw (0.0cm,-0.1cm) -- (0.0cm,0.4cm) ;
\node[anchor=south] at (0.0cm,\baselineskip) {$2^0$};
\draw (0.7\textwidth,-0.1cm) -- (0.7\textwidth,0.3cm) ;
\node[anchor=south] at (0.7\textwidth,\baselineskip) {\texttt{lpb}};
\draw (0.40\textwidth,-0.1cm) -- (0.40\textwidth,0.3cm) ;
\node[anchor=south] at (0.40\textwidth,\baselineskip) {\texttt{lim}};

\draw[decorate,decoration={brace,mirror,amplitude=0.3cm},thick]
(0,0) --
node[anchor=north,below=0.2cm]{product tree}
(0.40\textwidth,0);
\draw[decorate,decoration={brace,mirror,amplitude=0.3cm},thick]
(0.40\textwidth,0) --
node[anchor=north,below=0.2cm]{\begin{tabular}{c} ECM \\ extracted \end{tabular}}
(0.7\textwidth,0);
\end{tikzpicture}

\caption{Bucket sieve and product-tree sieve.
\label{fig:bucket-sieve}
\label{fig:product-tree-sieve}}
\end{figure}

A rough approximation of the memory required by the above procedure is
as follows, with $\#\cA$ denoting the size of the sieve area, and bounds
$2^{\texttt I}$ and \texttt{lim} being the two ends of the bucket-sieved range, as
represented in Figure~\ref{fig:bucket-sieve}.
\begin{align*}
    \text{memory required} &\approx
    \#\cA\times\sum_{\stackfrac{\fkp\in\cF_0}
{\fkp\ \text{bucket-sieved}}}\frac 1{\left|\Norm\fkp\right|}.\\
    &\approx
    \#\cA\times\left(\log\log\texttt{lim}-\log\log 2^{\texttt{I}}\right).
\end{align*}
    The formula above shows that if bucket sieving is used as
    described above for prime ideals around $2^{\texttt{I}}$, which is not very
    large,
the number of updates to store before applying
them becomes a burden. To alleviate this, and deal with (comparatively)
low-memory hardware, \cadonfs\ can be instructed to do the ``fill
buckets'' step above in several stages. Medium-size prime ideals (below a
bound called \texttt{bkthresh1}, mentioned in Figure~\ref{fig:bucket-sieve})
are actually considered for only up to 256 buckets at a time. Updates for
prime ideals above \texttt{bkthresh1}, on the other hand, are handled
in two passes. This
    leads to:%
\begin{align*}
    \text{memory required} &\approx
    \#\cA\times\left(\log\log\texttt{lim}-\log\log\texttt{bkthresh1}\right)\\
    &+\frac{\#\cA}{256}\left(\log\log\texttt{bkthresh1}-\log\log 2^{\texttt{I}})\right).
\end{align*}

\subsubsection{Interaction with batch smoothness detection.}
If the algorithms inspired by~\cite{djb-sf-2002} are used, the impact on
distribution and parallelization must be considered. The cost analysis
assumes that the product of the primes to be extracted has roughly the
same size as the product of \emph{survivors} to be tested. Only then can
we claim a quasi-linear asymptotic complexity. In this context,
a survivor is an $(a,b)$ pair for which the sieving done
on one side reveals a smooth or promising enough norm so that the
norm on
the other side will enter the batch smoothness detection.
The situation depends
on
the number of survivors per special-\fkq.

In the DLP-240 case, the sieving parameters are chosen to reduce the size
of the matrix. This has the consequence that the desired relations that are ``high
quality'' relations are rare, so that the number of survivors per
special-\fkq\ is small (about 7000 per special-\fkq, for $\#\cA=2^{31}$). In this setting, it makes sense
to accumulate all the survivors corresponding to all the
special-\fkq\ of a work unit in memory, and handle them at the end. There are so few
survivors that the cost of the batch smoothness detection remains small.
This strategy deviates from the asymptotic analysis but works well
enough, and does not interfere with the 
distribution techniques used by \cadonfs.

In the RSA-240 case, the situation is quite different. The number of
survivors per special-\fkq\ is high, so that the relative cost of the batch
smoothness detection is non-negligible. It is therefore important to
accumulate the correct number of survivors before starting to build the
product trees. In our setting, this corresponds to between 100 and 200
special-{\fkq}s, depending on their size. In order to keep the
implementation robust to these variations and to a non-predefined work unit size, 
we had the sieving software print the survivors to files.
A new file is started after a given number of survivors have been
printed. This way, the processing can be handled asynchronously by other
independent jobs. Again with simplicity and robustness in mind, we
preferred to have the production and the processing of the survivors
running on the same node, so as to avoid network transfers. Therefore the
survivor files were stored on a local disk (or even on a RAM-disk for
disk-less nodes). The next question is how to share the resources
on a single node, taking into account the fact that the top level of the product tree
involves large integer multiplications, which do not parallelize
efficiently, and yet consume a large amount of memory. After running some
experiments, we found that nodes with at least 4GB of RAM per physical
core could smoothly and efficiently accommodate the following setting:
\begin{itemize}
    \item One main job does the sieving on one side and
        continuously produces survivor files, each of them containing
        about 16M survivors. It uses as many threads as
        the number of logical cores on the node. In the parallelization
        strategy mentioned on page~\pageref{subsubsec:parallelization},
        only half of the RAM is considered
        available.
    \item About half a dozen parallel jobs wait for
        survivor files to be ready, and then run the batch
        smoothness detection followed by the final steps required to
        write relations. Each of these jobs has an upper limit of (for example) 8
        threads. The parallelization allows us to treat product trees and
        ECM curves in parallel, but each multiplication is
        single-threaded.
\end{itemize}
We rely on the task scheduler of the operating system to take care of the
competing jobs: in our setting the total number of threads that could in
principle be used is larger than the number of logical cores.
But since the jobs that process the survivors
are restricted to just one thread when performing a long multiplication,
it is important that the sieving makes full use of the cores during
these potentially long periods of time.

\subsection{Choosing parameters}

There are so many parameters controlling the relation collection, each of which can
be tuned and that interact in complex ways, that it is tempting to
choose them according to previous work and educated guesses based on what is
known to be efficient for smaller sizes where many full experiments can be
done.  However, techniques like batch smooth detection might only be
relevant for large enough sizes. We attempted as much as possible to be
rigorous in our parameter selection, and for this we developed dedicated
tools on top of \cadonfs\ to analyze a given set of parameters. First, 
we carried out sample sieving over a wide \fkq-range to deduce the
projected yield, with duplicate relations removed on the fly. Second,
 we developed a simulator that could infer the corresponding matrix size with good
accuracy, given some of the sample relations generated above. Both tools are detailed in §\ref{sec:simulation}.

Equipped with these tools, there was still a wide range of parameters to
explore. We give some general ideas about how we narrowed our focus to a subset of
the parameter ranges. This is different for the case of DL and IF.

For RSA-240, we first looked for appropriate parameters in the classical
setting where we sieve on both sides and can then allow as many large
primes on each side as we wish. It quickly became clear that a sieving
bound of $2^{31}$ was perhaps not optimal but close to it. Since
$2^{31}$ is also the current implementation limit of \cadonfs, this
settled the question.
Then the sieve area \cA\ has to be at least around this
size, so that sieving is amortized. The range of special-\fkq{}s is then
taken around the sieving bound.
We
can use the following rules
to choose good large prime bounds.
When doubling the large prime bounds, the number of
required relations roughly doubles. Therefore the number of (unique)
relations per special-\fkq\ should more than double to compensate for
this increased need. When this stops to be the case, we are around the
optimal value for the large prime bounds. These considerations gave us a first
set of parameters. We estimated that we were not too far from the
optimal. Then we
explored around it using our tools, also adding the possibility of having
a different choice of parameters to allow batch smoothness detection for
the large special-\fkq.

In the DLP-240 case, the choices of the sieving and large prime bounds
were dictated by constraints on the size of the resulting matrix.
The
general idea we used for this is that when the relations include at
most 2 large primes on each side, the matrix size after filtering
depends mostly on the sieving bound, which bounds the number of dense
columns in the matrix that will be hard to eliminate during the
filtering. Keeping the large prime bound small was also important to
reduce the number of survivors that enter the batch smoothness detection.
We used another empirical rule that helped us look for appropriate
parameters. In the case of composite special-\fkq\ where the number of
duplicate relations could quickly increase, keeping a \fkq-range whose
upper bound is twice the lower bound is a safe way to ensure that
the duplicate rate stays under control. In order to have enough \fkq{}s
in the range, the consequence of this was then to have them beyond the
large prime bound, which might look surprising at first.  Our simulation
tools were crucial to validate these unusual choices before running the
large-scale computation.

\subsection{Computational data}

%
\subsubsection{RSA-240.}

The relation collection for RSA-240 was carried out with large prime bounds of
36 bits on side~0 (the ``rational side'': $f_0$ is the linear polynomial) and
37 bits on side~1 (the ``algebraic side'': $f_1$ has degree~6).
We
used two parameter sets,
both with sieve area size $\#\cA = 2^{32}$.
We considered special-\fkq{}s on side~1.

For special-\fkq{}s whose norm is within 0.8G-2.1G,
we sieved on
both sides, with sieving bounds
$\texttt{lim}_0=\textrm{1.8G}$
and $\texttt{lim}_1=\textrm{2.1G}$.
We permitted cofactors no larger than $72$ bits on side~0 and $111$ bits on side~1, 
which allowed for up to 
two $36$-bit large primes on side 0 and
three $37$-bit large primes on side 1.

For special-\fkq{}s whose norm is within 2.1G-7.4G, we sieved only on
side~1 using  $\texttt{lim}_1=\textrm{2.1G}$ as above,
and used ``batch smoothness detection'' on side~0, as done in
\cite{EC:KDLPS17} (albeit the other way around). We allowed one fewer large
prime than above on side~1, which accounts for the contribution of the 
special-\fkq\
(see
Figure~\ref{fig:spq}).
The ``classical sieving'' took 280 core-years,
while ``sieving + batch smoothness detection'' took 514 core-years,
for a total of 794 core-years. 

\subsubsection{DLP-240.}

For DLP-240, we allowed
large primes of up to 35 bits on both sides.
We used composite special-\fkq{}s on side~0, within 150G-300G,
and prime factors between $p_{\text{min}} = 8192$ and 
$p_{\text{max}} = 10^8$ (see also Figure~\ref{fig:spq}).
Since $p_{\text{max}} < 150{\text G}$ and $p_{\text{min}}^3 >
300{\text G}$ this forced special-\fkq{}s with two factors. We
had 3.67G of these.

The sieve area size was $\#\cA = 2^{31}$. We sieved on side~0 only,
with a sieving bound of 
$2^{29}$. We used batch smoothness detection on
side~1 to detect ideals of norm below $2^{28}$. Up to two $35$-bit large
primes were allowed on each side.
This relation collection task took a total of 2400 core-years.

\section{Filtering}
\label{sec:filtering}
Prior to entering the linear algebra phase, the preprocessing step called
\emph{filtering} is mostly hampered by the size of the data set
that it is dealing with. Overall this phase takes negligible time, but it needs a
significant amount of core memory, and the quality of the filtering
matters. As filtering proceeds, the number of rows of the matrix
decreases, and its density (number of non-zero elements per row)
increases slightly.  Filtering can be stopped at any point, which we
choose to minimize the cost of the linear algebra computation that
follows.

Filtering starts with ``singleton'' and ``clique''
removal~\cite{Cavallar00}.
This reduces the
\emph{excess} (difference between the number of relations and the number
of ideals appearing in them) to almost zero.
Then follows another step, called \emph{merge} in \cadonfs,
which does some preliminary steps of Gaussian elimination
to reduce the matrix size, while increasing its density as little as possible.

Section~\ref{sec:simulation} describes simulations that were performed before the real filtering,
to estimate the final matrix size.
%
%
Filtering was done on a dual-socket, 56-core
Intel Xeon E7-4850 machine with
1.5TB of main memory. 
The \emph{merge} step was performed with the parallel algorithm from
\cite{bouillaguet:hal-02098114}.

\subsubsection{RSA-240.}

For special-\fkq{}s in 0.8G-7.4G,
we collected a total of 8.9G relations
which gave 6.0G unique relations.  See Table~\ref{tab:main} for exact figures.
After singleton removal (note that the initial excess was negative,
with 6.3G ideals), we had 2.6G relations on
2.4G ideals.
After ``clique removal'', 1.2G relations remained, with an excess of
160 relations.
The \emph{merge} step took about 110 minutes of wall-clock time,
plus 40 minutes of I/O time.
It produced a 282M-dimensional matrix with 200 non-zero
elements per row on average. 
We forgot to include around 4M free relations, which would have
decreased the matrix size by 0.1\%.

\subsubsection{DLP-240.}

For special-\fkq{}s in 150G-300G,
we collected a total of 3.8G relations, which gave 2.4G
unique relations. 
After singleton removal, we had 1.3G relations on 1.0G
ideals, and therefore had an enormous excess of around 30\%.
After ``clique removal'', 150M relations remained, with an excess of
3 more relations than ideals,
so we reduced the matrix size by a factor of almost 9.
The \emph{merge} step took less than 20 minutes, plus 8 minutes of I/O.
It produced a 36M-dimensional matrix, with an average of 253 non-zero
elements per row.
We generated several other matrices, with target density ranging from 200 to 275, 
but the overall expected time for linear algebra did not vary
much between these matrices.

\section{Linear algebra}
\label{sec:linalg}
We used the block Wiedemann
algorithm~\cite{Coppersmith94} for linear algebra.
In the short description below,
we let $M$ be the sparse matrix that defines
the linear system, and 
$\#M$ 
denotes its number of rows and columns.
We choose two integers $m$ and $n$ called
\emph{blocking factors}. We also choose
$x$ and $y$, which are blocks of $m$ and $n$ vectors. The main ingredient
in the block Wiedemann algorithm is the sequence
of $m\times n$ matrices $(x^TM^iy)_{i\geq0}$. We look at these matrices
column-wise, and form \emph{several} sequences, $n$ in the
DLP case, 
and $n/64$ in the factoring case, since for the latter it is worthwhile to handle
64 columns at a time because the base field is \GF 2.
The
algorithm proceeds as follows.
\begin{itemize}
    \item The \emph{Krylov} step computes the sequences. For each of the $n$
        sequences, this involves $(1/m+1/n)\cdot\#M$ matrix-times-vector
        operations. In the factoring case, the basic operation is
        the multiplication of $M$ by a \emph{block} of 64 binary
        vectors, in a single-instruction, multiple-data manner. 
        Note that sequences can be computed independently.
    \item The \emph{Lingen} step computes a matrix linear generator~\cite{BeLa94,DBLP:journals/jsc/Thome02,DBLP:conf/issac/GiorgiL14}.
    \item The \emph{Mksol} step ``makes the solution'' from the
        previously computed data. This requires $1/n\cdot\#M$ matrix-times-vector
        operations \cite[§7]{Kaltofen95}.
\end{itemize}

\subsection{Main aspects of the block Wiedemann steps}

In order to make good use of a computer cluster for the block Wiedemann
algorithm, several aspects must be considered. These are intended to
serve as a guide to the choice of blocking factors $m$ and $n$, along with
other implementation choices. To simplify the exposition, we assume below that
the ratio $m/n$ is constant. It is fairly typical to have $m=2n$.

First, the matrix-times-vector operation often \emph{must} be done on
several machines in parallel, as the matrix itself does not fit in RAM.
Furthermore, vectors that must be kept around also have a significant
memory footprint.  This opens up a wealth of MPI-level and thread-level
parallelization opportunities, which are supported by the \cadonfs\
software. Two optimization criteria matter: the time
per iteration, and the aggregated time over all nodes considered. Since
the computation pattern is very regular, it is easy to obtain projected
timings from a small number of matrix-times-vector iterations. Note that
the scaling is not perfect here: while having a larger number of nodes
participating in matrix-times-vector operation usually decreases the time
per operation, the decrease is not linear, since it is impeded by the
communication cost.

Since no communication is needed between sequences in the Krylov step,
it is tempting to increase the parameter $n$ in order to use more
coarse-grained parallelism.  If $n$ increases 
(together with $m$, since we assumed constant $m/n$),
the aggregate Krylov time over all nodes does not change, but
the time to completion does. In other words, the scaling is perfect. On the other hand, large blocking factors
impact the complexity of the Lingen step.
It is therefore important to predict
the time and memory usage of the Lingen step.

The input data of the Lingen step consists of $(m+n)\#M$ elements of the base
field (either \GF p or \GF 2). The space complexity of the quasi-linear
algorithms described
in~\cite{BeLa94,DBLP:journals/jsc/Thome02,DBLP:conf/issac/GiorgiL14} is
linear in
the input size, with an effectively
computable ratio. Their main operations
are multiplications and middle products of matrices
with large polynomial entries \cite{DBLP:journals/aaecc/HanrotQZ04}.
This calls for FFT transform caching: for
example, in order to multiply two $n\times n$ matrices, one can compute
transforms for
$2n^2$ entries, compute $n^3$ pointwise products and accumulate them to
$n^2$ transforms, which are finally converted back to entries of the
resulting product. However
this technique must be used with care. As described above, it needs
memory for $3n^2$ transforms. With no change in the running time, mindful
scheduling of the allocation and deallocation of transforms leads to only
$n^2+n+1$ transforms that are needed in memory, and it is fairly clear that it is
possible to trade a leaner memory footprint with a moderately larger run
time.

Another property of the Lingen step is that its most expensive operations
(multiplications and middle products of matrices
with large polynomial entries) parallelize well over many nodes and
cores.
This aspect was key to
the success of the block Wiedemann computation in the reported
computations.

The Mksol step represents only a fraction of the computational cost of
the Krylov step. It is also straightforward to distribute, provided that
some intermediate checkpoints are saved in the Krylov step.
In order to allow $K$-way distribution of the Mksol step, it is
sufficient to store checkpoints every $\#M/(nK)$ iterations during the
Krylov step, for a total storage cost of $Kn\cdot \#M$ base field elements,
typically stored on disk.

\subsection{Choosing parameters}

In line with the observations above, we used the following roadmap in order
to find appropriate parameters for linear algebra.
\begin{itemize}
%
    \item Run sample timings of the matrix-times-vector iterations
        with a variety of possible choices that the implementation offers:
        number of nodes participating in iterations, number of
        threads per node, and binding of threads to CPU cores.
        While it is possible to identify sensible choices via some rules of thumb, 
        the experience varies significantly with the
        hardware. We chose to pursue a simple-minded exploratory approach over an
        overly complex automated approach, whose benefit was
        unclear to us.
%
    \item
        Estimate the
        running time of the Lingen step with a simulated run.
        All internal steps of the algorithm used for the computation of
        the linear generator are well-identified tasks. Their individual
        cost is often tiny, but they must be repeated a large number of times.
        For example, for DLP-240 this involved $2^{18}$ repeated multiplications of
        polynomials of degree $1.1\times 10^6$ over \GF p.
        Obtaining a reasonable estimate of the timings is
        therefore fairly straightforward, although it is made somewhat more complex
        by including multithreading, parallelism at the
        node level, and memory constraints.
%
    \item 
        Estimate timings for the Mksol step, using
        techniques similar to the Krylov step.
        The wall-clock time also
        depends on the number of checkpoints that are saved during the
        Krylov step, as it governs the distribution of the work.

    \item
        The collected data gives expected wall-clock time
        and aggregated CPU time for the different steps as functions of
        the parameter choices. Then the only remaining step was to choose an optimum.
        Ultimately, the optimal choice very much depends on criteria visible
        only to an end user, or that are platform-specific.
        For example we had to take into
        account fixed compute budgets on one of the clusters that we
        used, as well as limits on the number of different jobs that can
        run simultaneously.
\end{itemize}

\subsection{Checkpoints}

Checkpoints have a number of uses in the computation. They allowed 
us to parallelize the Mksol step, to recover from failures, and in addition, they allow
offline verification of the
computation. All of the checks described in~\cite{DBLP:conf/issac/DumasKTV16}
are implemented in \cadonfs. This helped us diagnose problems with the
intermediary data that were caused by transient storage array failures.

As it turned out, there were mishaps during the linear algebra computation,
because some data files were affected by transient errors from the storage
servers, which thus affected the resulting computations on them. The ability to
verify the data offline more than saved our day.

\subsection{Computational data}

\subsubsection{RSA-240.}

We ran the block Wiedemann algorithm with parameters $m=512$, $n=256$.
The Krylov step used \emph{best-effort} jobs, using $n/64=4$
sequences, 8 nodes per sequence,
with two Intel Xeon Gold 6130 processors on each node and 64 virtual cores per node. The nodes were connected with Intel Omni-Path hardware. The
cost per matrix-times-vector product was 1.3 seconds, roughly 30\% of
which was spent in communications. This cost 69 core-years in total,
and the computation took 37 days of wall-clock time.  Despite the
best-effort mode, we were able to use the (otherwise busy) cluster more
than 66\% of the time.
The Lingen step was run on 16 similar nodes, and
took 13 hours (0.8 core-year).
The Mksol step was divided into 34
independent 8-node jobs, and took 13 core-years.
%
%

\subsubsection{DLP-240.}
We ran the block Wiedemann algorithm with parameters $m=48$, $n=16$. The
Krylov step used 4 nodes per sequence, with two Intel Xeon Platinum 8168
processors (96 virtual cores per node). The nodes were connected with
Mellanox EDR hardware. The cost per matrix-times-vector product was about
2.4 seconds, roughly 35\% of which was spent on communication. We used
$16\times4=64$ nodes almost full time for 100 days, for an aggregated
cost of 700 core-years.
The
Lingen step was run on 36 nodes, and took 62 hours (12 core-years).
The Mksol step was
divided into 70 independent 8-node jobs running simultaneously, and was
completed in slightly more than one day (70 core-years).
Note that these timings were obtained on slightly different hardware than
used elsewhere in this document.
Table~\ref{tab:main} reports our measurements with
respect to the Xeon Gold 6130
processors that we used as a reference, leading to a slightly smaller
aggregate cost (close to 650 core-years).

%

\section{Final steps: square root and descent}
\label{sec:descent+sqrt}

In the factoring context,
from the combination found by linear algebra we have a congruence of
the form $x^2\equiv y^2\mod N$, but we only know $x^2$, not $x$.
By computing two square roots we can write the potential factorization
$(x-y)(x+y)\equiv0\mod N$. This square root computation can be done with
Montgomery's square root algorithm~\cite{Montgomery94}, but a simple
approach based on $p$-adic lifting also works and has quasi-linear
complexity~\cite{DBLP:conf/waifi/Thome12}.

In the discrete logarithm context, the output of linear algebra consists
of a large database of known logarithms. To answer a query for the
logarithm of an arbitrary element, the \emph{descent} procedure needs to
search for relations that establish the link between the queried element
and the known logarithms. This requires a specially adapted version of the
relation collection software.

\subsubsection{RSA-240.} 

The final computations for RSA-240 were performed on the same hardware
that was
used for the filtering step (\textsection\ref{sec:filtering}).
%
%
After reading the 1.2G relations that survived the clique removal,
and taking the quadratic characters into account,
we obtained 21 dependencies
in a little more than one hour. 

For the square root step, we used the direct (lifting) approach described
in
\cite{DBLP:conf/waifi/Thome12}.
Each dependency had about 588M relations.
On the rational side, we simply multiplied the corresponding $F_0(a,b)$ values,
which gave an integer of about 11Gb; we then computed its square root using
the Gnu MP library \cite{GMP612}, and reduced it modulo $N$.
As usual, we were eager to get the factors as soon as possible.
This square root step therefore prompted some development effort to have
a program with a high degree of parallelism.
The rational square roots of the first four
dependencies were obtained on November 22 in the end of the afternoon;
each one took less than two hours of wall-clock time, with a peak memory
of 116Gb.
The first algebraic square root was computed in 17 hours wall-clock time,
and finished at 02:38am on November 24.  Further code improvements
reduced the wall-clock time to only 5 hours. We were a bit lucky, since
this first square root led to the 
factors of RSA-240.

\subsubsection{DLP-240.} 

The individual logarithm step (called ``descent'' for short) is
dominated by the first \emph{smoothing} step, after which
subsequent descent trees must be built. We followed a
practical strategy similar to the one described in~\cite{EC:FGHT17}, but
since there is no rational side, sieving is not available during the
smoothing step. Therefore, for a target element $z$, we tried many
randomizations $z'=z^e$ for random exponents $e$, each of them being
lifted to the $f_1$-side, LLL-reduced to get a small representative
modulo the prime ideal above $p$ used in the diagram of
Figure~\ref{fig:nfs-diag}, and then tested for smoothness with a chain of ECMs.
In fact, we handled a pool of candidates simultaneously, keeping only the
most promising ones along the chain of ECM they go through.
This strategy, which has been implemented in {\cadonfs}, can be viewed
as a practical version of the admissibility strategy described
in~\cite[Chapter~4]{BarbulescuPhD}, which yields the best known complexity for this
step. The chain of ECMs that we used is tuned to be able to extract with a
high probability all prime factors up to 75 bits. But of course,
many non-promising candidates are discarded early in the chain.
We enter the descent step when we find a
candidate which is $100$-bit smooth.

The descent step itself consists of rewriting all prime ideals of
unknown logarithm in terms of ideals of smaller norms, so that we can
ultimately deduce the discrete logarithms from the ones that were
computed after the linear algebra phase. As predicted by the theory,
these descent trees take a short amount of time compared to the
smoothing.
Although this last step can be handled automatically by the general
\cadonfs\ machinery, we used some custom (and less robust) tools written
to avoid lengthy I/O, and to reduce the wall clock time because,
then again, we were eager to get the result.

No effort was made to optimize the CPU-time of this step which took a few
thousand core-hours, mostly taken by the smoothing phase that was run
on thousands of cores in parallel.  (This resulted in us finding several smooth
elements, while only one was necessary).

\begin{table}[t]
    \begin{center}
  \begin{tabular}{ccc}
    & RSA-240 & DLP-240 \\ \hline
    \emph{polynomial selection} & 76 core-years & 152 core-years \\
    $\deg f_0,\deg f_1$           & 1,6 & 3,4            \\ \hline
    \emph{relation collection}
      \\
      large prime bounds & $\texttt{lpb}_0=2^{36}$, $\texttt{lpb}_1=2^{37}$        & $\texttt{lpb}_0=\texttt{lpb}_1=2^{35}$\\
      type of special-\fkq{}s&  prime (side 1) & composite ($\fkq_1\fkq_2$, side 0)\\
    method               & a: lattice sieving for $f_0$ and $f_1$,                 & lattice sieving for $f_0$ \\
                         & $\left|\Norm\fkq\right|\in [0.8\textrm G,2.1\textrm G]$ & and factorization tree for $f_1$ \\ 
                         & b: lattice sieving for $f_1$ and                        & $\left|\Norm\fkq\right| \in [150\textrm G,300\textrm G]$ \\
                         & factorization tree for $f_0$, & \\
                         & $\left|\Norm\fkq\right| \in [2.1\textrm G,7.4\textrm G]$ & \\
    sieving bounds       &  a: $\texttt{lim}_0 = 1.8\textrm G$, $\texttt{lim}_1=2.1\textrm G$            & $\texttt{lim}_0 \approx 540M$ \\ 
                         &b: $\texttt{lim}_1 = 2.1\textrm G$\\
    product tree bound
      & b: $\texttt{lim}_0 = 2^{31}$
      & $\texttt{lim}_1=2^{28}$\\
      \# large primes per side &  a: $(2,3)$, b: $(2,2)$ & $(2,2)$\\
    sieve area \cA       & $2^{32}$         & $2^{31}$ \\
    raw relations        & $8\,936\,812\,502$& 3\,824\,340\,698 \\
    unique relations     & $6\,011\,911\,051$ & 2\,380\,725\,637 \\
    total time           & 794 core-years & 2400 core-years \\
    \hline
      \emph{filtering}  \\
    after singleton removal& $2\,603\,459\,110 \times 2\,383\,461\,671$
                         & $1\,304\,822\,186 \times 1\,000\,258\,769$ \\
    after clique removal & $1\,175\,353\,278 \times 1\,175\,353\,118$
                         & $149\,898\,095 \times 149\,898\,092$ \\
    after merge          & 282M rows, density 200
                         & 36M rows, density 253 \\ \hline
    \emph{linear algebra}\\
    blocking factors& 
      $m=512$, $n=256$ &
      $m=48$, $n=16$ \\
      Krylov & $4\times 8$ nodes, 69 core-years
      & $16\times 4$ nodes, 544 core-years\\
      Lingen & 16 nodes, 0.8 core-years & 36 nodes, 12
      core-years\\
      Mksol
      & $34\times 8$ nodes, 13 core-years
      & $70\times 8$ nodes, 69 core-years\\
    total time           & 83 core-years & 625 core-years \\
      \hline
  \end{tabular}

  \caption{Comparison of 795-bit factoring and computing 795-bit prime field
    discrete logarithm. ``$x\times y$ nodes'' means that $x$ independent jobs, each
    using $y$ nodes simultaneously, were run, either in parallel (most
    often) or
    sequentially (at times). All timings are scaled to physical cores of Intel Xeon Gold
    6130 processors.\label{tab:main}}
    \end{center}
\end{table}

\section{NFS Simulation}
\label{sec:simulation}

The goal of an NFS simulation is, given a set of parameters,
to predict the running times of the main phases of the algorithm
together with relevant data like the size of the matrix. In this section, we give
some details about the tools that we developed and used before running
the computations.

We assume that we are given the number $N$ to factor (or the prime $p$, in
the DLP case), together with a pair of polynomials, maybe not the
final optimal choice, but reasonably close to the best we expect
to find. We also have a set of NFS parameters that we want to test.

The general idea is to let the sieving program run for a few
special-\fkq{}s and use the resulting relations as models,
to  produce at very high speed \emph{fake} relations corresponding to the full
range of special-\fkq. Then the filtering programs are applied to these
relations, in order to produce a fake matrix. By timing a few
matrix-times-vector operations, we can also estimate the linear algebra
cost.

\paragraph{Sampling relations.} For a set of special-\fkq{}s evenly
sampled in the target special-\fkq\ range, we run the sieving program with
the target parameters and keep the corresponding relations for future
use. The relations that would be found as duplicates in the real
filtering step are removed. These can be detected quickly as follows: for
each prime ideal in the factorization of the relation that belongs to the
special-\fkq\ range and is less than the current special-\fkq, we analyze
whether the relation would have been found when sieving for this smaller
special-\fkq. In \cadonfs, this online duplicate removal option
has almost no impact on the sieving time and is almost perfect.

\paragraph{Producing fake relations.} Let $I$ be a special-\fkq\
ideal for which we want to produce fake relations. We start by looking at
a set $S_I$ of special-{\fkq}s of about the same size that were
sieved during the sampling phase. The number of fake relations that will
be produced for $I$ is chosen by picking a random element $I'$ in
$S_I$ and taking as many relations  as for $I'$.
Then, for each relation to be produced, we pick a random relation $R$
among all the relations of all the special-\fkq{}s in $S_I$ and modify
it:
replace the special-\fkq\ by $I$,
and replace each of the other
ideals by another one picked randomly among the ideals of
norm $\pm20\%$ of the original norm on the same side.
We therefore keep the general statistical properties of
the relations (distribution of the number of large primes on each side,
weight of the columns taking into account the special-\fkq{}s, etc.).

\paragraph{Emulating the filtering.} The filtering step can be run
as if the relations were genuine. The only difference is that the
duplicate removal must be skipped, since our relation set is based on a
sample in which duplicate relations have been removed. This produces a
matrix whose characteristics resemble the ones of the true matrix, and
which can be used to anticipate the cost of
linear
algebra.

\paragraph{A mini-filter approach.} The simulation technique we have
sketched takes a tiny fraction of the total time of the real computation.
However, in terms of disk and memory space, it has the same requirements
and this might be prohibitive when exploring many different parameters.
We propose a strategy to faithfully simulate the whole computation with
all the data being reduced by a \emph{shrink factor}, denoted $\sigma$.
Typical values will be between 10 and 100 depending on the size of the
experiment that has to be simulated and the expected precision.

For each special-\fkq, the number of fake relations we produce is divided by
$\sigma$. If this number is close to or smaller than 1, this is done in a
probabilistic way; for instance if we have to produce $0.2$ relations for
the current special-\fkq, then we produce one relation with probability
$0.2$ and zero otherwise. This reduces the number of relations (rows of
the matrix before filtering) by a factor $\sigma$, as expected.  In order
to also reduce the number of columns, we divide the index of each ideal
in the relation by $\sigma$, keeping each side independent of the other one.
This simultaneous shrinking of rows and columns has the following
properties:
\begin{itemize}
    \item The average weight per row and per column is preserved (not
        divided by $\sigma$);
    \item More generally the row- and column-weight distributions are
        preserved;
    \item The effects of the special-\fkq{}s are preserved: the average
        weight of the columns corresponding to special-\fkq{}s will be
        increased by the average number of relations per special-\fkq\ of
        that size;
    \item The variations of the weight distribution of the columns
        that depend on their size and on whether they are above or below
        the sieving bound are preserved.
\end{itemize}
The filtering step can then be applied to these shrunk fake relations.
The final matrix is expected to be $\sigma$ times smaller than the true
matrix. Of course, this matrix cannot be used to directly estimate the
running time of the linear algebra step: it needs to be expanded again.
But often, already being able to compare the size of the matrices can
help discard some bad parameter choices before only a few of the
most promising ones can be simulated again, perhaps with a smaller value of
$\sigma$ or no shrink at all.

This very simple and easy-to-implement technique produces good results as long as
the shrink factor $\sigma$ is not too large. In our still very partial
experiments, if the final shrunk matrix has at least a few tens of
thousands of rows and columns, then the result is meaningful.

\paragraph{Estimates for RSA-240 and DLP-240.}
For DLP-240, we used such a set of fake relations in August 2018,
i.e., at the very start of the relation collection.
The set contained 2244M unique relations.  After singleton
and clique removal we obtained a matrix of size 144M, and after
merge we obtained a matrix of size 37.1M with average density 200.

The closest run with real relations was carried out in the end of February 2019,
with 2298M unique relations, giving a matrix of size 159M after singleton 
and clique removal, and a matrix of size
about 40.7M with average density 200 after merge. In comparison, our actual
computation collected a few more relations, and would have led to a
matrix with 38.9M rows if we had stopped the filtering at density 200, while
we ultimately chose to use the matrix with density 250 instead, which had
36.2M rows.
Hence our technique allowed us to obtain
an early precise estimate---with error below 10\%---for the final
matrix size.

For RSA-240 we also used the \emph{mini-filter} approach, with a
\emph{shrink factor} of 100.
In December 2018 we started with a matrix of size 66M,
which gave a matrix of size
15.7M after singleton and clique removal, and a matrix of size 3.3M 
after merge. This is within 17\% of the size
of the real matrix we obtained in mid-2019,
taking into account the shrink factor of 100.

\paragraph{Prospects for more precise simulations.}
Our simulation machinery is still experimental, but allowed us to be sure
beforehand that we would be able to run the linear algebra with our
available resources. This was especially relevant for DLP-240, where
the sieving parameters are chosen with the aim of reducing the size
of the matrix.

A more systematic study is needed to validate this simulator and improve
its precision. In particular, we expect better results from taking a more sophisticated
 strategy for building fake relations based on a sample of real
ones. Also, with the shrink factor, a better handling of the columns
corresponding to small very dense ideals would probably help,
with if possible a different algorithm for discrete logarithm and integer
factorization.

\section{Conclusion}
\label{sec:conclusion}

It is natural to ask how our computational records compare to
previous ones, and how much of our achievement can be attributed to
hardware progress.  A comparison with RSA-768, which was factored 10
years before the present work, would have very limited meaning.
Instead, we prefer to compare to the DLP-768 record from 2017. Extrapolations
based on the $L(1/3,c)$ formula suggest that DLP-240 is about 2.25 times
harder than DLP-768.  The article~\cite{EC:KDLPS17} reports that the DLP-768
computation required 5300 (physical) core-years on Intel Xeon E5-2660
processors, and further details indicate that the relation collection
time was about 4000 core-years.  We ran the \cadonfs\ relation collection
code with our parameters on exactly identical processors that we happened
to have available. The outcome is that on these processors, we would have
been able to complete the DLP-240 relation collection in only about 3100
core-years. 
So our parameter choice allowed us to do more work in less time.

The timeline of previous records is misleading: RSA-768 was factored in
2009, and DLP-768 was solved in 2016. Furthermore, the latter required
more resources than the former (a raw ratio of core-years gives a factor of
3.5, but this would be amplified significantly if we used identical
hardware). This contributes to the idea that for similar size problems,
finite field discrete logarithm is much harder than integer factoring.
Our experiments show that this difference is not as striking as commonly
thought. Based on the data in Table~\ref{tab:main}, the ratio is only 3.3
with identical hardware: 3177 versus 953 core-years. Furthermore, this
ratio only holds if we consider the DLP modulo ``safe primes'', which
leads to more difficult linear algebra. In the so-called ``DSA-like'' situation
where we seek discrete logarithms in a small subgroup of $\GF p^*$,
the linear algebra becomes easier, which leads to trade-offs between
relation collection and linear algebra: the ratio
is likely to drop, perhaps close to or maybe even below a factor of two.

Another reason that finite field discrete
logarithm is considered to be much harder than integer factoring is that the linear
algebra step is believed to be a major bottleneck. It is true to some extent:
in our computation, as well as in previous ones, the balance in
aggregated CPU time is shifted towards less expensive linear algebra, because
more infrastructure (in particular, interconnect and storage) is required for linear algebra than for sieving. 
However, it is important to notice that with
adequate parameter choices, large sparse linear systems occurring in NFS
computations \emph{can} be handled, and at this point we are not facing a
technology barrier.

\iffinal
\paragraph{Acknowledgements.}
We thank Gérald Monard and the support team of the EXPLOR computing center
for their help, the engineers of the Grid'5000 platform, and Joshua Fried, Luke Valenta, and Rafi Rubin for sysadmin help at the University of Pennsylvania.

\paragraph{Funding.}
This work was possible thanks to a 32M-hour allocation on the Juwels
super-computer from the PRACE research infrastructure.
Experiments presented in this paper were carried out using the Grid'5000 testbed, supported by a scientific interest group hosted by Inria and including CNRS, RENATER and several Universities as well as other organizations (see \url{https://www.grid5000.fr}).
This work was supported by the French ``Ministère de l'Enseignement Supérieur
et de la Recherche'', by the ``Conseil Régional de Lorraine'', by the
European Union, through the ``Cyber-Entreprises'' project, and by the US National Science Foundation under grant no.~1651344.
High Performance Computing resources were partially provided by the EXPLOR centre hosted by the University de Lorraine.
Computations carried out at the University of Pennsylvania were performed on Cisco UCS servers donated by Cisco.
\fi

\makeatletter
\def\@doi#1{\href{http://dx.doi.org/#1}{doi:\texttt{#1}}\catcode`\_8}
\def\doi{\catcode`\_11\relax\@doi}
\makeatother
\bibliographystyle{splncs04}
\bibliography{cryptobib-extract,paper}

\appendix
\section{Challenge results}
\label{apx:dlp}

To prove that we have computed discrete logarithms modulo
$p=\text{RSA-240} + 49204$, we consider the integer $y$ whose
hexadecimal expansion corresponds to the ASCII encoding of the sentence
``The magic words are still Squeamish Ossifrage'' (without newline, and
with big-endian convention, i.e., $y=\texttt{0x54...65}$). In $\GF p^*$,
the discrete logarithm of $y$ to base $g=5$ is
\begin{eqnarray*}
    \log_5 y &=&
\scriptstyle
    926031359281441953630949553317328555029610991914376116167294\\&&
\scriptstyle
    204758987445623653667881005480990720934875482587528029233264\\&&
\scriptstyle
    473672441500961216292648092075981950622133668898591866811269\\&&
\scriptstyle
    28982506005127728321426751244111412371767375547225045851716
\end{eqnarray*}
\label{apx:factors}%
\label{apx:factors-pm1}%
With respect to RSA-240, the factors are given by $\text{RSA-240}=p\times q$, with
\begin{eqnarray*}
\scriptstyle
    \text{RSA-240} &=&
\scriptstyle
    124620366781718784065835044608106590434820374651678805754818\\&&
\scriptstyle
    788883289666801188210855036039570272508747509864768438458621\\&&
\scriptstyle
    054865537970253930571891217684318286362846948405301614416430\\&&
\scriptstyle
    468066875699415246993185704183030512549594371372159029236099,\\
    p&=&
\scriptstyle
    509435952285839914555051023580843714132648382024111473186660\\&&
\scriptstyle
    296521821206469746700620316443478873837606252372049619334517,\\
    q&=&
\scriptstyle
    244624208838318150567813139024002896653802092578931401452041\\&&
\scriptstyle
    221336558477095178155258218897735030590669041302045908071447.
\end{eqnarray*}
\section{RSA-250 details}
We selected the following polynomial pair, with ${\rm
Res}(f_0,f_1)=48\times\text{RSA-250}$:
{\small
\begin{eqnarray*}
  f_1 &=& {86130508464000}\, x^6 
    - {81583513076429048837733781438376984122961112000}\,  \\
    &-& {66689953322631501408}\, x^5 
    - {1721614429538740120011760034829385792019395}\, x \\
    &-& {52733221034966333966198}\, x^4 
    - {3113627253613202265126907420550648326}\, x^2 \\
    &+& {46262124564021437136744523465879}\, x^3 \\
  f_0 &=& {185112968818638292881913}\, x 
    - {3256571715934047438664355774734330386901}\, 
\end{eqnarray*}
}
We used the following
important parameters: $\texttt{lim}_{0,1} = 2^{31},\
\texttt{lpb}_0=2^{36},\ \texttt{lpb}_1=2^{37}$. We used lattice sieving for $f_0$ and $f_1$ when
$\left|\Norm\fkq\right|\in [1\textrm G,4\textrm G]$ with $2$ large primes
for $f_0$ and $3$ large primes for $f_1$,
and lattice sieving for
$f_1$ and factorization tree for $f_0$ when
$\left|\Norm\fkq\right|\in [4\textrm G,12\textrm G]$, with $2$ large primes
for both $f_0$ and $f_1$.
The sieve area was $2^{33}$.

Sieving gave a total of 8\,745\,268\,073 raw relations, of which
6.1G were unique (70.1\%).
After the singleton removal, there were 2.7G relations remaining on
2.6G ideals. After clique removal, there were 1.8G
relations remaining, 
with an excess of 160.
The merge step produced a matrix of about 405M rows, with average density
252 (about 100G non-zero elements).
We computed 64 dependencies with the block Wiedemann algorithm, with
parameters $m=1024$ and $n=512$.
For each dependency, the square root step took about 2.3 hours on the
rational side (on a dual-socket, 56-core Intel Xeon E7-4850), and 10.5
hours on the algebraic side.

We obtained the factorization $\text{RSA-250}=p\times q$, with
\begin{eqnarray*}
\scriptstyle
  \text{RSA-250} &=&
\scriptstyle
    214032465024074496126442307283933356300861471514475501779775492\\&&
\scriptstyle
    088141802344714013664334551909580467961099285187247091458768739\\&&
\scriptstyle
    626192155736304745477052080511905649310668769159001975940569345\\&&
\scriptstyle
    7452230589325976697471681738069364894699871578494975937497937,\\
    p&=&
\scriptstyle
    641352894770715802787901901705773890848250147429434472081168596\\&&
\scriptstyle
    32024532344630238623598752668347708737661925585694639798853367,\\
    q&=&
\scriptstyle
    333720275949781565562260106053551142279407603447675546667845209\\&&
\scriptstyle
    87023841729210037080257448673296881877565718986258036932062711
\end{eqnarray*}
Using the same reference (Intel Xeon Gold 6130 at 2.10GHz) as
elsewhere in this paper,
the total computation time for RSA-250 was roughly 2700 core-years,
including 2450 core-years for the sieving step and 250 core-years for
the linear algebra step.

Complete details of the RSA-240, DLP-240, and RSA-250 computations can be
found in
\begin{center}
\url{https://gitlab.inria.fr/cado-nfs/records}
\end{center}

\end{document}